\documentclass[11pt]{article}
\pdfoutput=1
\usepackage{amsfonts}
\usepackage{graphicx}
\usepackage{amsmath,bm}
\usepackage{amssymb}
\usepackage{latexsym}
\usepackage{mathrsfs}
\usepackage{color}
\usepackage{slashed}
\usepackage[nosort]{cite}
\usepackage{booktabs}
\usepackage[font=small]{caption}
\usepackage{multirow}
\usepackage{soul} %
\usepackage{setspace} %
\makeatletter
\def\section{\@startsection {section}{1}{\z@}{-3.5ex plus -1ex minus
 -.2ex}{2.3ex plus .2ex}{\large\bf}}
\def\subsection{\@startsection{subsection}{2}{\z@}{-3.25ex plus -1ex
minus -.2ex}{1.5ex plus .2ex}{\normalsize\bf}}
\makeatother
\makeatletter

\@addtoreset{equation}{section}

\makeatother

\def\bea{\begin{eqnarray}} \def\eea{\end{eqnarray}}
\def\be{\begin{equation}} \def\ee{\end{equation}} 
\def\nn{\nonumber}
\usepackage[colorlinks,linkcolor=black,citecolor=blue,urlcolor=blue,linktocpage]{hyperref}

\begin{document}
\setcounter{page}{0}
\thispagestyle{empty}

\parskip 3pt

\font\mini=cmr10 at 2pt

\begin{titlepage}
~\vspace{1cm}
\begin{center}

{\LARGE \bf $\lambda \phi^4$ Theory I: The Symmetric Phase
\\[8pt] Beyond NNNNNNNNLO}

\vspace{0.6cm}

{\large
Marco~Serone$^{a,b}$, Gabriele~Spada$^a$, and Giovanni~Villadoro$^{b}$}
\\
\vspace{.6cm}
{\normalsize { \sl $^{a}$ 
SISSA International School for Advanced Studies and INFN Trieste, \\
Via Bonomea 265, 34136, Trieste, Italy }}

\vspace{.3cm}
{\normalsize { \sl $^{b}$ Abdus Salam International Centre for Theoretical Physics, \\
Strada Costiera 11, 34151, Trieste, Italy}}

\end{center}
\vspace{.8cm}
\begin{abstract}

Perturbation theory of a large class of scalar field theories in $d<4$  can be shown 
to be Borel resummable using arguments based on Lefschetz thimbles.
As an example we study in detail the  $\lambda \phi^4$ theory in two dimensions in the 
$Z_2$ symmetric phase.
We extend the results for the perturbative expansion of several quantities up to N$^8$LO 
and show how  the behavior of the theory 
at strong coupling can be recovered successfully using known resummation techniques. 
In particular, we compute the vacuum energy and the mass gap for values of the coupling up to 
the critical point, where the theory becomes gapless and lies in the same universality class 
of the 2d Ising model. Several properties of the critical point are determined 
and agree with known exact expressions. 
The results are in very good agreement (and with
comparable precision) with those obtained by other non-perturbative approaches, such as
lattice simulations and Hamiltonian truncation methods.  

\end{abstract}

\end{titlepage}

\tableofcontents

\section{Introduction}

There has been considerable interest in recent years on the possible use of the Lefschetz thimble decomposition of integrals and path integrals (see e.g.\ ref.\cite{Witten:2010cx} for a physics oriented review) 
to better understand the properties of perturbative expansions in quantum mechanics and quantum field theory (see e.g.\ ref.\cite{Aniceto:2018bis} for a recent review in the context of 
the theory of resurgence).
In particular, a connection between the Lefschetz thimble decomposition of path integrals and Borel summability of perturbative series has been recently pointed out in the context of one-dimensional quantum mechanical systems with bound-state potentials and discrete spectra \cite{Serone:2017nmd,Serone:2016qog}. More specifically, it has been shown that the saddle-point expansion 
of a Lefschetz thimble (i.e. the steepest descent manifold of a saddle point) is Borel resummable. It has also been shown how to properly define perturbative expansions that are Borel resummable to the exact result
in theories, such as the quantum mechanical symmetric double well, where ordinary perturbation theory is not.

The aim of this paper (and of its companion \cite{Z2Broken}) is to begin an extension of the above results in quantum field theory (QFT).
We start in section~\ref{BorelGeom} by showing that arbitrary $n$-point correlation functions (Schwinger functions) are Borel reconstructable from their loopwise expansion in a broad class of Euclidean 2d and 3d scalar field theories.
These includes basically all UV complete scalar field theories on $\mathbb{R}^d$ with the exception of those with continuously connected degenerate vacua in 2d. 

The Borel summability of Schwinger functions in this class of theories is deduced by a change of variables in the path integral and the absence of non-trivial 
positive finite actions solutions to the classical equations of motion. This is equivalent to establishing that the original path integral over real field configurations coincides with a single Lefschetz thimble.
The ordinary loopwise expansion around the vacuum $\phi=0$ can then be interpreted as the saddle point expansion over a Lefschetz thimble, hence Borel summability is guaranteed.
The Borel resummed Schwinger functions coincide with the exact ones in a given phase of the theory. If phase transitions occur at some finite values of the couplings, the analysis is more subtle. The analytically continued Schwinger functions may not coincide with the physical ones in the other phase. 
We discuss these subtle issues for the two-dimensional $\phi^4$ theory in subsection~\ref{subsec:BSPTT}, where we point out that Schwinger functions  that are analytically continued from the unbroken to the broken phase correspond to expectation values of operators over a vacuum violating cluster decomposition. In the rest of the paper we focus on the 2d $\phi^4$ theory, that we study in some detail.

The Lagrangian density of the $\phi^4$ theory reads
\be
{\cal L} = \frac 12 (\partial_\mu \phi)^2 + \frac 12 m^2 \phi^2 + \lambda \phi^4\,.
\label{Lagr}
\ee
The theory is super-renormalizable. Only the $0$-point function
and the momentum independent part of the $2$-point functions 
are superficially divergent.
The coupling constant $\lambda$ is finite and there is no need of a field wave function renormalization, while counterterms are required for the vacuum energy and the mass term.
The effective expansion parameter of the theory is the dimensionless coupling
\be
g \equiv \frac{\lambda}{m^2}\,.
\ee
Using certain bounds and analytic properties of the Schwinger functions, the perturbation series of arbitrary correlation functions had already been rigorously proved  to be Borel resummable,
though only for parametrically small coupling constant and large positive squared mass, i.e.\  for $g\ll 1$ \cite{Eckmann}.  

The RG flow of this theory is well-known. In the UV the theory becomes free, for any value of $g$ and $m^2$.
The flow of the theory in the IR depends on $g$ and $m^2$. For $m^2>0$ and $0\leq g< g_c$ the theory has a mass gap and  the $\mathbb Z_2$ symmetry $\phi\to-\phi$ is unbroken.
At a critical value of the coupling $g=g_c$ the theory develops a second-order phase transition \cite{Chang:1976ek,Simon:1973yz}.
At the critical point the theory is a conformal field theory (CFT) 
in the same universality class of the 2d Ising model. 
Above $g_c$ the theory is in the $\mathbb Z_2$ broken phase.

For $m^2<0$ and $0<g< \widetilde g_c$ the theory has a mass gap and the $\mathbb Z_2$ symmetry is broken.
At $g=\widetilde g_c$ the second-order phase transition occurs. The $\mathbb Z_2$ symmetry is restored for $\widetilde g_c<g<\widetilde g_c'$. At $g>\widetilde g'_c$, the theory returns to a broken phase. 
Interestingly enough, a simple duality can be established between the $\mathbb Z_2$ unbroken and broken phases that allows, among other things, to relate the three critical values $g_c$, $\widetilde g_c$ and $\widetilde g'_c$  \cite{Chang:1976ek}
(see also ref.\cite{Rychkov:2015vap} for a nice recent discussion). In this paper we will study the 2d $\phi^4$ theory when $m^2>0$ and $0\leq g\leq  g_c$, postponing to ref.~\cite{Z2Broken} the study of the $\mathbb Z_2$ broken phase.

In section~\ref{sec:pert-coeff} we compute the perturbative series expansion up to order $g^8$ of
the 0-point and 2-point Schwinger functions.
This requires the evaluation of Feynman diagrams up to nine and eight loops for the 0-point and the 2-point functions, respectively, that have been computed using 
numerical methods. We define, as usual, the physical mass $M$ as the simple pole of the Fourier transform
of the 2-point function (at complex values of the Euclidean momenta, corresponding to real Lorentzian momenta) and provide the perturbative series for $M^2$.
The two main results of this section are the perturbative expressions for the vacuum energy $\Lambda$ and $M^2$ in eqs.(\ref{LambdaFullSeries}) and (\ref{mphFullSeries}), respectively.

In section~\ref{sec:numerics} we describe the resummation procedures that we have adopted to estimate the exact Borel transform from their known truncated expressions: the Pad\'e-Borel
approximants and the conformal mapping method.  They have been already used with success in the study of the 2d $\phi^4$ theory since the earlier works of \cite{Baker:1976ff,Baker:1977hp,LeGuillou:1979ixc}.  
We explain in some detail the methods  we have followed, in such a way that  the interested reader might be able to reproduce our findings. 

\begin{figure}[t!]
\centering
              \includegraphics[width=78mm]{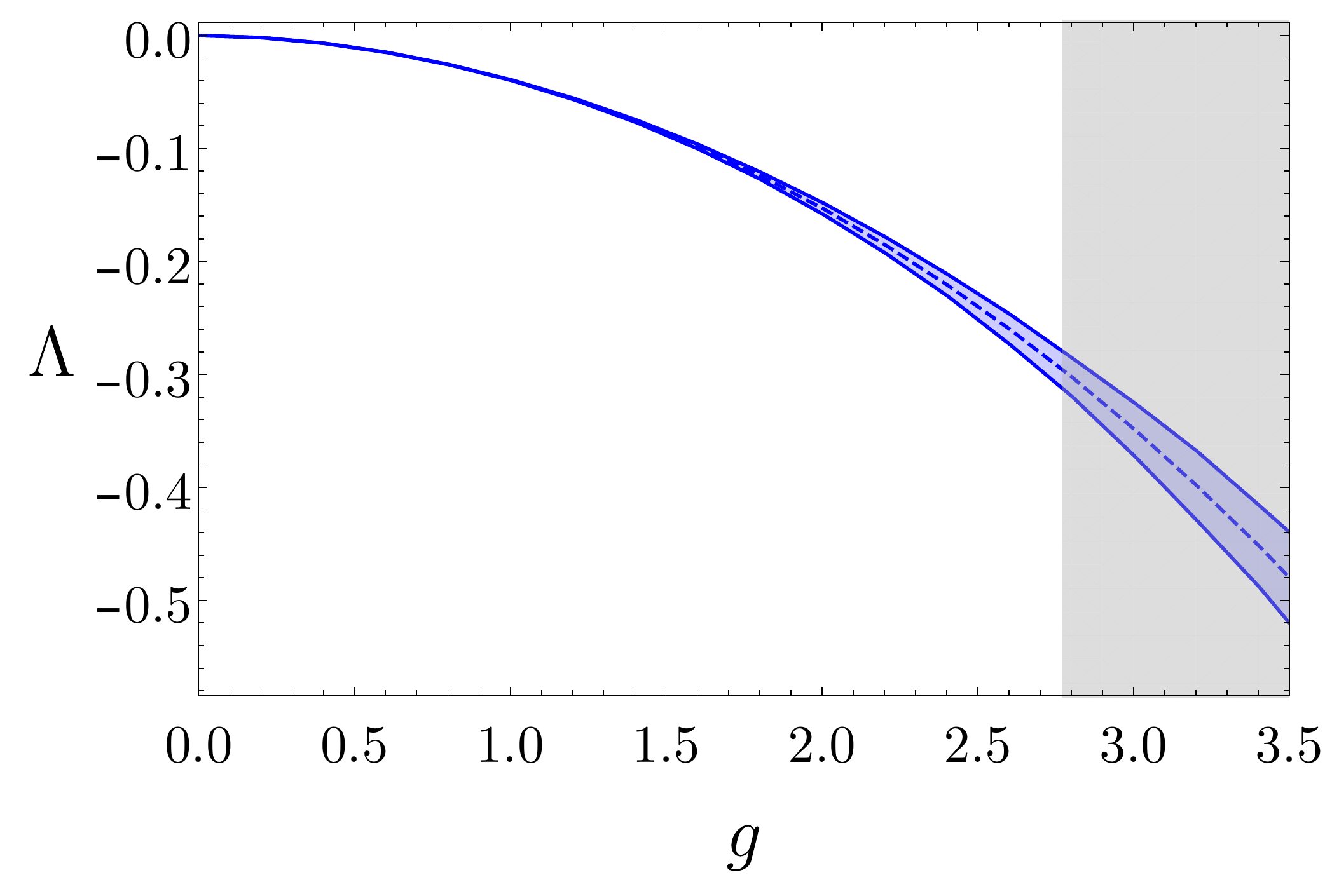}~~%
           \includegraphics[width=78mm]{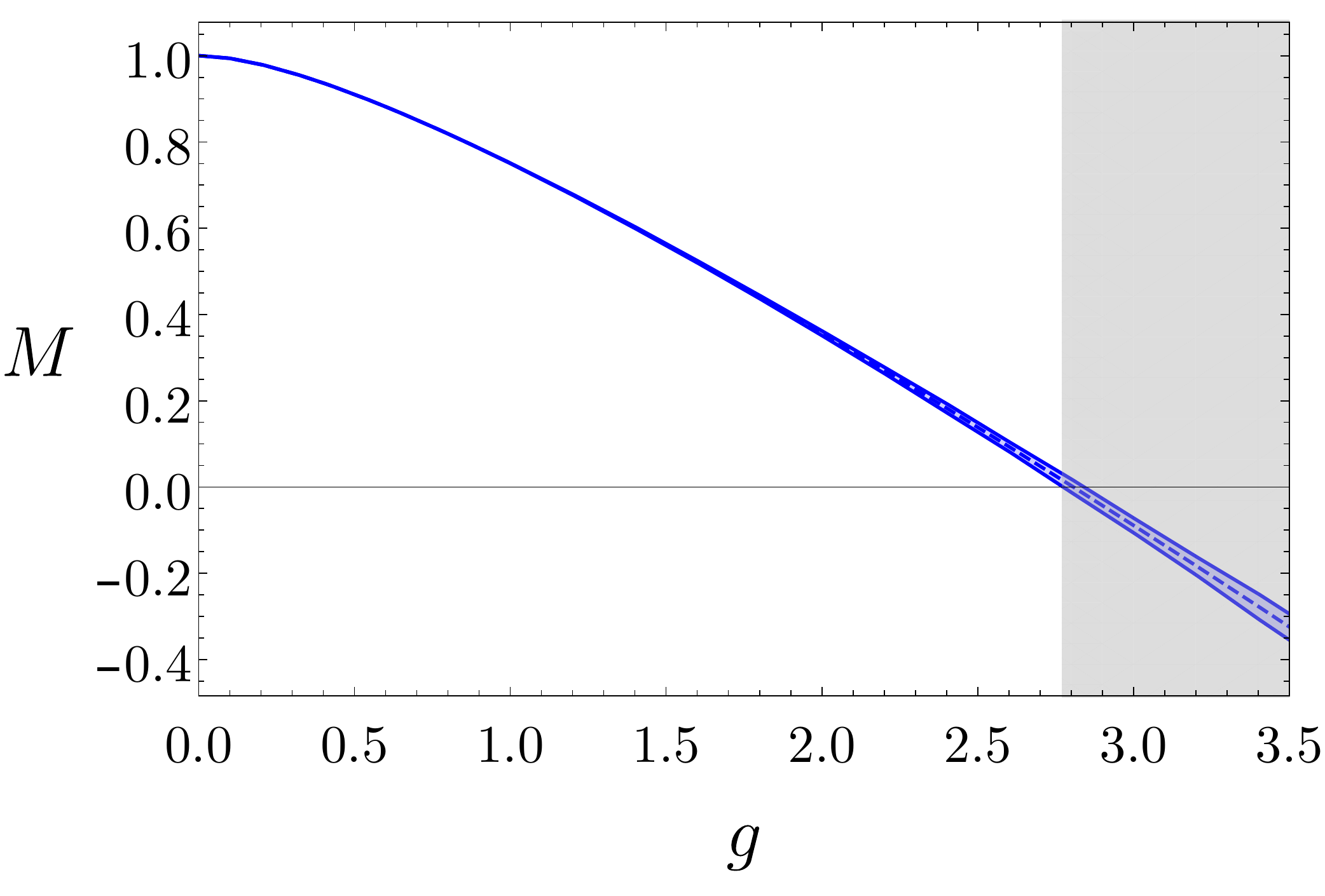}%
\caption{\label{fig:LambdaMass}
The vacuum energy $\Lambda$ (left) and the mass gap $M$  (right) as a function of the coupling constant $g$ obtained by Borel resumming the perturbative series using the coefficients up to the $g^8$ order. The values reported of $\Lambda$ and $M$ for $g>g_c\approx 2.8$, where a phase transition occurs, refer to the extrapolation using the Borel resummed function.}
\end{figure}

The final results of our Borel resummation procedures are reported in section~\ref{sec:results}. We determine $\Lambda$ and $M$ for any  $g$ up to $g\gtrsim g_c$.
The critical coupling $g_c$ is defined as $M(g_c)=0$. By resumming a proper function of $M$ we compute the critical exponent $\nu$, defined as
\be
M(g) \propto |g_c-g|^\nu , \quad g\rightarrow g_c \,.
\label{nuDef}
\ee
The exponent $\eta$ is extracted directly by its definition as the power-like decay of the two-point function at the critical point, where the theory is a CFT:
\be
\langle \phi(x) \phi(0) \rangle_{g=g_c} = \frac{\kappa}{|x|^\eta} \,.
\label{etaDef}
\ee
We also compute the non-perturbative renormalization constant $\kappa$ appearing in eq.~(\ref{etaDef}), required
to properly match the fundamental field $\phi$ with the 2d Ising magnetization operator $\sigma$.
For the convenience of the reader, we report in table \ref{FinalRes} the values of $g_c$, $\nu$,  $\eta$  and $\kappa$ and in fig.\ref{fig:LambdaMass} $\Lambda$ and $M$ as a function of $g$, as found in section~\ref{sec:results}. These are the most important numerical results of the paper. 

The strong coupling behavior of the 2d $\phi^4$ theory has been addressed by several other methods, including lattice Monte Carlo, lattice matrix product states and Hamiltonian truncations.
We compare our findings with those in the literature, particularly with refs.\cite{Elias-Miro:2017xxf,Elias-Miro:2017tup}, in section~\ref{sec:comparisons}, finding very good agreement overall.

\begin{table}[t!]
\centering
{
\renewcommand{\arraystretch}{0.8}
\begin{tabular}{cccc}
\toprule
$g_c$   &  $\nu$  &  $\eta$  &  $\kappa$  \\
\midrule
$2.807(34)$ & $0.96(6)$ & $0.244(28)$ & $0.29(2)$  \\
\bottomrule
\end{tabular}
}
\caption{The values of the critical coupling $g_c$, the critical exponents $\nu$, $\eta$ and the 2-point function normalization $\kappa$ at the critical point as defined in eq.~(\ref{etaDef}), found in this paper. 
The exact values of the exponents are $\nu=1$ and $\eta=1/4$ \cite{Onsager:1943jn}.} 
\label{FinalRes}
\end{table}

It is important to emphasize here, in the context of the 2d $\phi^4$ theory, the main differences and analogies between our resummation procedure and those already developed in the literature:
the $\epsilon$-expansion \cite{Wilson:1971dc} and the so called fixed dimension resummation \cite{Parisi:1993sp}.
The $\epsilon$-expansion allows some control on the critical theory,  thanks to the existence of weakly coupled fixed points for $\epsilon\ll1$.
Its validity is restricted to the critical theory and the Borel summability of the expansion is still conjectural \cite{Brezin:1976vw}.
The second method is closer in spirit to ours, and essentially boils down to a different renormalization scheme.
It has however been used only to study the theory at criticality and to determine the critical exponents of the theory by means of the same methods used in the $\epsilon$-expansion,
namely finding zeros of a suitably defined $\beta$-function. 
In this paper, aside from giving theoretical evidence for the Borel summability of Schwinger functions,
resummation techniques are used for the first time to study the $\phi^4$ theory away from criticality.
Moreover, the critical coupling and the critical exponents $\nu$ and $\eta$ are extracted in a different way, using the two-point function only, 
without relying on the above mentioned $\beta$-function, which requires the (more demanding) computation of a four-point function.
We compare our results with those obtained using the above mentioned resummation methods in section \ref{subsec:comparisonresum}.
We have computed the zero momentum two-point function, its derivative and the four-point function,  required to compute the $\beta$-function and $\eta$ 
in the scheme of ref.\cite{Parisi:1993sp}, extending by 2 orders the known series in $\beta$ and by 3 orders the one in $\eta$ \cite{Orlov:2000wn}.

We conclude in section~\ref{sec:conclusions}. Some details on the perturbative series expansions are reported in appendix~\ref{sec:appendix}.

\section{Borel Summability in $d<4$ Scalar Field Theories  }
\label{BorelGeom}

In this section we show how the Borel summability of Schwinger functions in the $\lambda \phi^4$ theory can be more easily inferred and extended to a large class of scalar field theories (though in a less rigorous way than refs.\cite{Eckmann,Magnen})  using a geometric approach borrowed by Picard-Lefschetz theory. 

We start by considering the Euclidean path integral 
\begin{equation}
F= \int {\cal D}\phi\, G[\phi]\, e^{-S[\phi]/\hslash} \,, 
\end{equation}
where we momentarily reintroduced $\hslash$, $S[\phi]$ is the Euclidean action
\begin{equation}
S[\phi]=\int d^dx\, \Bigl[\frac12 (\partial \phi)^2+V(\phi) \Bigr]\,,
\label{Sren}
\end{equation}
with $V(\phi)$ a generic polynomial potential bounded from below and $G[\phi]$ an arbitrary polynomial function of fields, 
product of local operators at different space-time points, corresponding to $n$-point Schwinger functions in the theory.
In order to avoid the danger of having an ill-defined path integral $V(\phi)$ should not contain irrelevant couplings; we restrict to the case of super-renomalizable potentials to
avoid complications from possible renormalons~\cite{tHooft}. Hence in $d=2$ $V(\phi)$
is a generic polynomial function bounded from below,\footnote{Most likely this is a conservative assumption and our considerations extend to more
general potentials.} while in $d=3$ it is a polynomial of degree up to four.  
For simplicity, we have omitted to write the space dependencies of $G$ and of the resulting Schwinger functions $F$. 
A proper definition of $F$ in general requires renormalization. 
The counterterms  needed to reabsorb the  divergences  
are subleading in a saddle point expansion in $\hslash$ and can be reabsorbed in the factor $G$, which clearly is no longer a polynomial in the fields.\footnote{For simplicity, we are assuming here that the saddle point expansion is well-defined for any field mode, namely ${\rm det}\, S''\neq 0$. 
Whenever this condition is not met, the corresponding zero mode should be evaluated exactly.} 
The counterterms do not change the saddle point structure of $S[\phi]/\hslash$, as long as the convergence of the path integral at large field values is dictated by $S$,
condition automatically satisfied in our case by the absence of marginal  and irrelevant couplings.

Without lack of generality we can choose $\phi_0=0$ for the value of the absolute minimum
of the classical potential (for the moment we assume that it is unique, we will discuss the case of
degenerate minima afterwards) and normalize the path integral so that $S[0]=0$. 
Out of the infinite integration variables of the path integral 
we now single out the one corresponding to the value of the action, namely we rewrite
our path integral as \cite{tHooft,howls,Gukov:2016njj,Serone:2017nmd}
\begin{equation} \label{eq:S2t-trick}
F= \int_0^\infty \!\! dt\,e^{-t} 
\int {\cal D}\phi\, G[\phi]\, \delta\bigl (t-S[\phi]/\hslash\bigr)=\int_0^\infty \!\! dt\,e^{-t} {\cal B}(\hslash t)\,,
\end{equation}
\emph{viz.} an integral over all possible values of the action $t=S/\hslash$ of the function 
${\cal B}(\hslash t)$, which is a path integral restricted to the section of phase space with fixed
action.\footnote{The fact that ${\cal B}(\hslash t)$ is indeed a function of the product $\hslash t$ was
shown explicitly in ref.\cite{Serone:2017nmd}.} Note that the function ${\cal B}$ in eq.~(\ref{eq:S2t-trick}) is the Laplace transform of $F$ and as such
it corresponds to its Borel transform when expanded in $\hslash$. The manipulation above is legit as long as the change of variables $t=S[\phi]/\hslash$ is non-singular, i.e.\ as long as $S'[\phi]\neq0$. Hence eq.~(\ref{eq:S2t-trick}) holds as long as there are no finite action critical points (aside from the trivial one $\phi_0=0$) for real field configurations. 
Following ref.~\cite{Brezin:1976wa}, a generalization of Derrick's theorem \cite{Derrick:1964ww} can be used to show that the action defined in eq.~(\ref{Sren}) does not have any non-trivial critical points with finite action. 
Analogously to the quantum mechanical case discussed in ref.\cite{Serone:2017nmd}, the combination of reality and boundedness
of the action and the presence of a unique critical point makes the domain of integration
of our path integral a single Lefschetz thimble away from Stokes lines (i.e. it does not intersect other saddle points). In this way Borel summability of the perturbative series  to the exact result  is guaranteed \cite{Serone:2017nmd}.\footnote{Derrick's argument about the absence of non-trivial critical points in scalar theories formally work also for non-integer dimensions. As such it could in principle be used as above to show the Borel summability of the $\epsilon$-expansion. However, this reasoning would require a non-perturbative definition of the path integral of theories with non-integer dimensions, which is unknown to us.} 
The above derivation is easily generalizable for multiple scalar fields.

If the minimum of the classical potential is not the global one, we generally have more than one finite action critical point (e.g.\ bounce solutions), the domain of integration
of the path integral is on a Stokes line and correspondingly the perturbative series will not be Borel resummable.

When the global minimum is not unique due to spontaneous symmetry breaking, by definition the action has multiple finite action critical points.\footnote{We do not discuss here the subtle case of 
degenerate vacua not related by symmetries.} At finite volume, this would again imply  that the domain of integration of the path integral is on a Stokes line and the perturbative series is non Borel resummable. 
In the infinite volume limit, however, all such vacua are disconnected from each other and the path integral should be taken in such a way that only one of such vacua is selected.
This request is crucial to guarantee that correlation functions satisfy the cluster decomposition property. The vacuum selection is achieved by adding a small explicit symmetry breaking term in the action that removes the degeneracy, taking the infinite volume limit, and then removing the extra term (more on this in the next subsection). In the infinite volume limit we are now in the same situation as before with a single global minimum.
The generalized Derrick's theorem forbids the existence of finite action solutions. The loopwise expansion around the selected vacuum $\phi^i=\phi_0^i$ can be interpreted as the saddle point expansion over a Lefschetz thimble and Borel summability is guaranteed. This argument does not apply to scalar theories with a continuous family of degenerate vacua in $d=2$,  because Derrick theorem does not hold in this case.\footnote{The connection between the absence of positive action critical points and the Borel summability of the perturbative series was proposed already in ref.~\cite{Brezin:1976wa}, which also noted the intriguing relation between Borel summability and the existence of spontaneous symmetry breaking. However, the arguments of ref.~\cite{Brezin:1976wa} could not establish that the resummed series reproduce the exact result, while the Lefschetz thimble perspective allows us to fill this gap and to put on firmer grounds the conjecture of ref.~\cite{Brezin:1976wa}.} In particular, our results imply that 2d and 3d $\phi^4$ theories with $m^2<0$, whose Borel resummability was not established before (see e.g.\ chapter 23.2 of ref.\cite{GlimmJaffe}), are in fact Borel resummable.

The discussion above applies in the absence of phase transitions. 
Whenever they occur at some finite values of the couplings, the analysis is more subtle because the infinite volume limit has to be taken with more care. 
We will discuss in some detail the case of interest in this paper, the $\phi^4$ theory in $d=2$ dimensions.
 
\subsection{Borel Summability and Phase Transitions in the $\phi^4$ Theory}
\label{subsec:BSPTT}

We consider here 
\be
V(\phi) = \frac 12 m^2 \phi^2 + \lambda \phi^4\,,
\label{VpotPhi4}
\ee
with $m^2>0$, $\lambda>0$ in $d=2$ dimensions. We can set $\hslash=1$, since the loopwise expansion is equivalent
to the expansion in $g=\lambda/m^2$.
For small enough $g$ the theory has a mass gap and the Schwinger functions are expected to be analytic in the coupling.
The situation becomes more delicate at and beyond the critical point $g\geq g_c$ where the theory undergoes a second order phase-transition~\cite{Glimm:1975tw,Chang:1976ek}. The latter typically suggests the presence of non-analyticities of $n$-point functions around the critical point.
On the other hand, the r.h.s.\ of eq.~(\ref{eq:S2t-trick}) could be well defined for any $g$, suggesting that the Schwinger functions could be smooth even beyond $g_c$.
Consider for example the 1-point function $\langle \phi \rangle$, order parameter
of the $\mathbb Z_2$ symmetry breaking. 
Since the path integral is an odd function of $\phi$,  $\langle \phi \rangle$ is obviously identically zero, and thus analytic and Borel summable, 
for any value of $g$.  For $g<g_c$, $\langle \phi \rangle=0$ is the correct result, but for $g>g_c$ the theory is in a $\mathbb Z_2$ broken phase and it fails to reproduce the ``right" value  of the order parameter $\pm v\neq 0$.

The apparent paradox is explained by recalling that vacuum selection in presence of spontaneous symmetry breaking requires the introduction 
of an explicit $\mathbb Z_2$ symmetry breaking term $\epsilon$ which should be turned off only {\it after} the infinite volume limit is taken.\footnote{In textbook perturbative treatments of spontaneous symmetry breaking  such procedure is not necessary because we select ``by hand" the vacuum around which to expand our fields according to the minima of the classical (or perturbative effective) potential.} 
Phase transitions show up as non-analiticities  of the correlation functions $F$ after the $V\rightarrow \infty$ and $\epsilon\rightarrow 0$ limits are taken.
We should consider the theory in a finite volume $V$ and modify the action as, for example, 
\begin{equation}
S_{V,\epsilon}[\phi]=S_V[\phi]+\epsilon \int_V d^2x\, \phi(x) \,,
\label{Smod}
\end{equation}
and define the correlation functions as 
\begin{equation}
F_{{\rm SSB}}\equiv \lim_{\epsilon\to 0}  \lim_{V\to \infty} F_\epsilon(V)\,,
\end{equation}
where, crucially, the limit $\epsilon\to0$ is taken after the infinite volume one.
In our arguments about Borel summability of scalar theories with a classically unique global minimum 
neither $V$ nor $\epsilon$ entered in the discussion, since we had $V=\infty$ and $\epsilon=0$ to start with. 
The Schwinger functions $F$ that are reconstructed by Borel resummation in this way
would correspond to take the limits in the opposite way:
\begin{equation}
F \equiv  \lim_{V\to \infty}  \lim_{\epsilon\to 0} F_\epsilon(V)\,.
\end{equation}
For $g\geq g_c$ we expect that  $F\neq F_{{\rm SSB}} $
as a  result of the non commutativity of the two limits.
In particular, for the one-point function we would have
\be\begin{split}
\langle \phi \rangle_{{\rm SSB}} =  & \; v  \lim_{\epsilon\to 0^\pm} \rm{sign}\,(\epsilon) \,,\\
\langle \phi \rangle = & \; 0\,,
\end{split}\ee
where $v$ is the non-perturbative value of the order parameter.

Resummation of the perturbative series for $F$ in the presence of an explicit breaking term is not straightforward (there is actually no need to consider finite volume, since this limit should be taken first)
and will not be done in this paper. 
Denoting by $|\pm\rangle$ the two vacua in the $\mathbb Z_2$ breaking phase, in the absence of the proper $\epsilon\to 0$ limit selecting either the vacuum $|+\rangle$ or $|-\rangle$,
for $g\geq g_c$ the vacuum is a linear combination of the form 
\be
|\alpha\rangle = \cos \alpha |+\rangle + \sin \alpha |-\rangle\,.
\label{etaVac}
\ee 
The condition $\langle \phi \rangle=0$ fixes $\alpha=\pi/4$ in eq.~(\ref{etaVac}), but it is useful to keep it generic  in what follows.
The Schwinger functions $F$ for $g\geq g_c$ correspond then to correlation functions around the vacuum $|\alpha\rangle$. As well-known, such vacua
violate cluster decomposition (see e.g.\ the end of section 19.1 of ref.\cite{weinberg}). 
This can easily be seen by considering for instance the large distance behavior of the connected two-point function $\langle \phi(x) \phi(0)\rangle_c$:
\be
\lim_{|x|\rightarrow \infty} \langle \alpha |  \phi(x) \phi(0)| \alpha \rangle_c = \sum_{n=\pm} \langle \alpha| \phi(0)| n\rangle \langle n | \phi(0) |\alpha \rangle -   \langle \alpha |  \phi(0)| \alpha \rangle^2 = \sin^2(2\alpha) v^2\,,
\label{ClusterDec}
\ee
which does not vanish for $\alpha=\pi/4$ (while it does for $\alpha =0, \pi/2$, as expected).

Summarizing,  Borel resummation of perturbation theory around the unbroken vacuum, 
when applied beyond the phase transition point and without explicit breaking terms,
reconstructs the correlation functions $F$  in the wrong vacuum where cluster decomposition is violated. 
 The correlation functions $F$ do not coincide with the ordinary ones,
$F_{{\rm SSB}}$, in the broken phase in any of the two vacua $|\pm\rangle$. 

We will mostly focus our attention on the simplest Schwinger functions, the 0- and the 2-pt functions. 
The presence of a phase transition can still be detected by looking at the mass $M(g)$ as a function of the coupling and defining $g_c$ as $M(g=g_c)=0$.
Since the vacua $|\pm\rangle$ are degenerate, the vacuum energy $\Lambda$, 
computed in the $|\alpha\rangle$ vacuum through resummation for $g\geq g_c$, 
 is expected to  coincide  with the vacuum energy in any of the two $\mathbb Z_2$ broken vacua. 
 
The interpretation of the resummed mass gap $|M|$ for $g\gtrsim g_c$ is more subtle
and we postpone the discussion of this region in ref.~\cite{Z2Broken}.  
In all the plots we present in the paper, the region with $g> g_c$ is shaded to highlight the $\mathbb Z_2$ unbroken phase $0\leq g \leq g_c$, where our computations should not present any subtleties.

\section{Perturbative Coefficients up to 9 Loops}
\label{sec:pert-coeff}

\begin{figure}[t]
\centering
\includegraphics[scale=1]{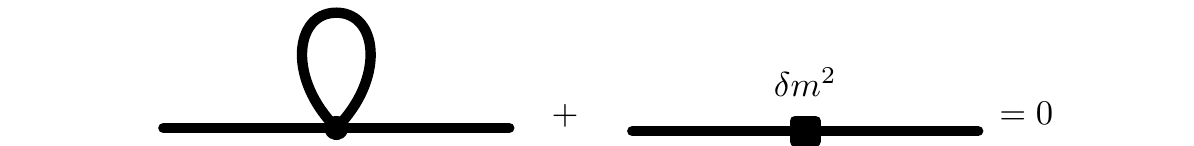}
\caption{The divergent one-loop diagram is exactly canceled by the mass counterterm. Within this scheme we can forget all the diagrams with lines that start and end at the same quartic vertex, their contribution being zero.}
\label{fig:normal-ord}
\end{figure}

The superficial degree of divergence  $\delta$ of a connected Feynman graph ${\cal G}$ in 2d scalar theories without derivative interactions is simply given by
\be
\delta({\cal G}) = 2 L - 2 I=2-2 V \,,
\label{SDD}
\ee
where $L$ is the number of loops, $I$ the number of internal lines, $V$ the number of
vertices and the last expression follows from $I=L+V-1$. The only superficially divergent 
graphs are therefore those with 0 and 1 vertex. The former correspond to the divergence of normalization of the free-theory path integral, while the latter correspond to diagrams
with loops from contraction of fields in the same vertex. It follows that 2d theories without
derivative interactions can be renormalized simply by normal ordering.
In particular in the $\phi^4$ theory superficial divergences only occur in the $0$-point function from $V=0$ and $V=1$ graphs (the 2-loop ``8''-shaped graph),
and in the momentum independent part of the $2$-point function with $V=1$ (1-loop  tadpole propagator graph).
Correspondingly, only the vacuum energy  and the mass term require the introduction of the counterterms 
$\delta \Lambda$ and $\delta m^2$. We choose $\delta m^2$ such that it completely cancels the contribution coming from the tadpole diagram (see fig.~\ref{fig:normal-ord}).
We choose $\delta \Lambda$ such that it  completely cancels the one and two-loop divergencies described above, so that $\Lambda = {\cal O}(g^2)$. The presence of these counterterms makes all $n$-point Schwinger functions finite to all orders in perturbation theory. In this scheme we do not need to compute the large set of diagrams involving tadpoles (their contribution in a different scheme can be reproduced by a proper shift of the mass). The scheme above coincides with the normal ordering, which is commonly used in other calculations of the $\phi^4$ theory.
This makes possible a direct comparison of scheme-dependent quantities such as the value of the critical coupling $g_c$  (see section \ref{sec:comparisons}).

In the following we focus on the $0$- and the $2$-point functions. 
Multi-loops computations present two obvious challenges: i) classifying the different topologically distinct graphs and computing their multiplicities, ii) evaluating the loop integrals.
In the $\phi^4$ theory point i) can be addressed with a number of tools (e.g.~FORM, FeynArts, QGRAF, \dots) but we find that it is possible to perform brute force Wick contractions with a simple Mathematica code up to the order $g^6$. In order to overcome memory limitations and reach the order $g^8$ we use the method described in ref.\cite{Kleinert:1999uv}. %
A non-trivial check of the topologies and multiplicities of the Feynman diagrams is performed in the one-dimensional $\phi^4$ theory by verifying that the integration of the diagrams reproduces the perturbative series 
of the quantum mechanical anharmonic oscillator, which can easily be computed to
very high orders with the recursion relations by Bender and Wu  \cite{Bender:1971gu,Bender:1990pd}. 

An analytic approach to point ii) is possible but quite challenging, so we opt for a numerical evaluation of the loop integrals. To this purpose it is convenient to write the Feynman diagrams 
in configuration, rather than momentum, space. The tree-level two-point function reads
\be
G_0(x) = \frac{1}{2\pi} K_0(m x)\,,
\label{eq:G0}
\ee
where $x = \sqrt{x_\mu x_\mu}$ and $K_n$ is the modified Bessel function.  For large $x$, $K_0(x)\propto  e^{-x}/\sqrt{x}$ and such exponential decay of $G_0$ makes a configuration space approach 
quite suitable to a numerical evaluation. We perform the integration of the diagrams in polar coordinates using the Montecarlo VEGAS algorithm \cite{Lepage:1977sw} as implemented in the {\tt python} module {\tt vegas}.
Since the number of different topologies rapidly increases with the order, calculating diagram by diagram becomes unpractical and summing up the diagrams seems a better approach. However, summing up many diagrams, and thus evaluating more complicated integrals, leads to a less precise estimate of the integral. For these reasons we find convenient to split the diagrams in bunches (ranging from 5 to 40 diagrams each) in such a way to have a balance between the number of evaluations and the final precision obtained. Typically the number of points used in the Montecarlo integration for each bunch is $10^9$ at order $g^8$, which roughly delivers a precision of $10^{-5}$ in the estimate of the integral. We checked the numerical convergence of the integrals when the number of points is increased.
Since order by order different diagrams contribute with the same sign the precision improves in the final sum.

\subsection{Vacuum Energy}

We compute the vacuum energy up to order $g^8$, which corresponds to evaluate one-particle irreducible (1PI) graphs with up to nine loops.
The number of topologically distinct graphs (in the chosen scheme) as a function of $g$ is reported in table \ref{tab:n-diag-0}.
For illustration, up to order $g^4$, these are
\bea
\Lambda & = &
 - 12 \,  \vcenter{\hbox{\includegraphics[scale=.33]{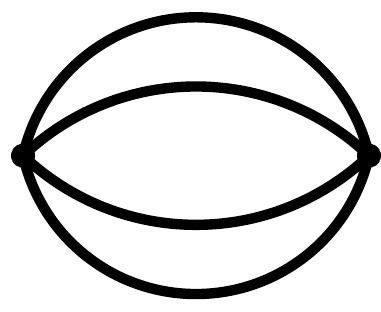}}} \, g^2
 + 288 \, \vcenter{\hbox{\includegraphics[scale=.33]{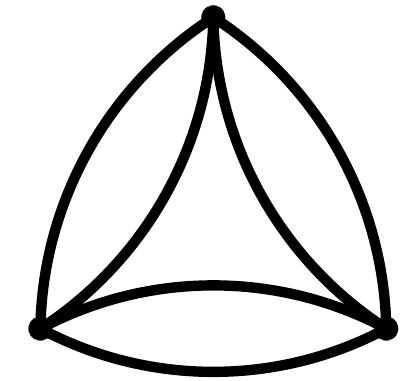}}} \, g^3
 + \\
 &&- \left( 2304 \, \vcenter{\hbox{\includegraphics[scale=.33]{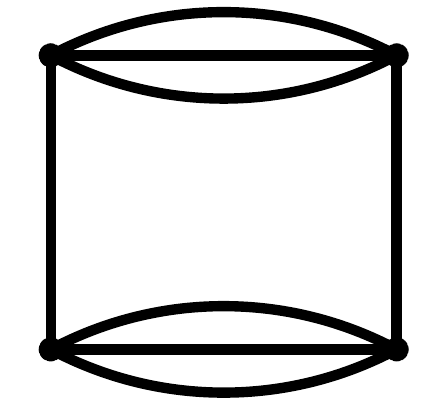}}}
 + 2592 \, \vcenter{\hbox{\includegraphics[scale=.33]{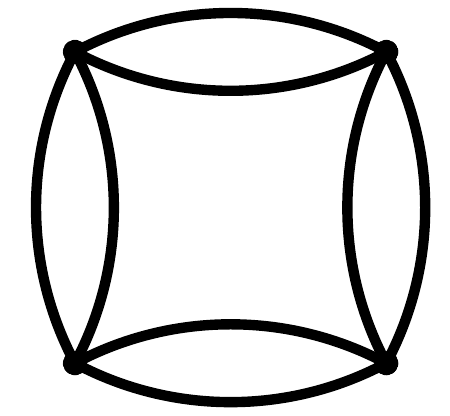}}}
 + 10368 \, \vcenter{\hbox{\includegraphics[scale=.33]{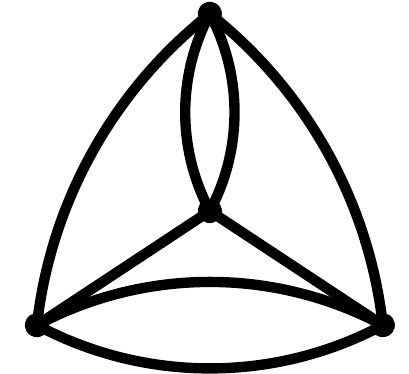}}} \right) g^4
 + {\cal O}(g^5)\,, \nn
 \eea
 where the numbers indicate the multiplicities of the different topologies.

  \begin{table}[t]
 \centering
 \begin{tabular}{cccccccc}
 \toprule
 $g$ & $g^2$ & $g^3$ & $g^4$ & $g^5$ & $g^6$ & $g^7$ & $g^8$ \\ 
 \midrule
 0 & 1 & 1 & 3 & 6 & 19 & 50 & 204 \\
 \bottomrule
 \end{tabular}
 \caption{Number of topologically distinct 1PI 0-pt diagrams without self-contractions.}
 \label{tab:n-diag-0}
 \end{table}

Evaluating all Feynman diagrams leads to our final expression for 
the perturbative expansion of $\Lambda$ up to $g^8$ order:
\begin{equation}
\begin{split}
\frac{\Lambda}{m^2} = 
&- \frac{21 \zeta(3)}{16 \pi^3} \, g^2 + \frac{27\zeta(3)}{8\pi^4} \, g^3 -0.116125964(91) \, g^4 + 0.3949534(18) \, g^5 \\ &- 1.629794(22) \, g^6 + 7.85404(21) \, g^7 - 43.1920(21) \,g^8 + {\cal O}(g^9)\,.
\end{split}
\label{LambdaFullSeries}
\end{equation}
The numbers in parenthesis indicate the error in the last two digits due to the numerical integration. The coefficients proportional to $g^2$  and $g^3$ are computed analytically.\footnote{We thank A.L. Fitzpatrick
for having pointed out to us that the $g^3$ coefficient can be analytically computed.}  

\subsection{Physical Mass}
The mass gap can be defined directly from the 2-pt function as
$M=-\lim_{x\to \infty}\log(\langle \phi(x) \phi(0) \rangle)/x$.
Equivalently, it can be computed from the smallest zero of the 1PI two-point function 
for complex values of the Euclidean momentum (corresponding 
to the real on-shell momentum in Minkowski space):
\be
\widetilde \Gamma_2(p^2=-M^2)\equiv 0\,,
\label{mphDef}
\ee
where 
\begin{equation}
 \widetilde \Gamma_2(p^2) = \int d^2 x \, e^{i p \cdot x } \, \Gamma_2(x) = p^2 + m^2 + {\cal O}(g^2)
 \end{equation}
 is the Fourier transform of the configuration space 1PI two-point function $\Gamma_2(x)$, that has no correction of ${\cal O}(g)$ in 
 the chosen scheme. We are interested to get an expansion of $M^2$ in powers of $g$ of the form
 \begin{equation}
 M^2 = m^2(1+ \sum_{k=2} a_k \, g^k) \equiv m^2 + \delta \,.
 \label{mphExp}
 \end{equation}  
 Plugging eq.~(\ref{mphExp}) in eq.~(\ref{mphDef}) gives
 \begin{equation}
 0 = \widetilde \Gamma_2 (-M^2)=\widetilde \Gamma_2 (-m^2 -\delta) =  \sum_{n=0}  \frac{\widetilde \Gamma_2^{(n)}(-m^2)}{n!} (-\delta)^n\,,
  \label{Gamma2m2}
 \end{equation}
where the Taylor coefficients can be computed directly as follows
 \be
 \widetilde \Gamma_2^{(n)}(-m^2) \equiv \left. \left[\frac{d}{dp^2}\right]^n\,\widetilde \Gamma_2 (p^2)\right |_{p^2=-m^2}=
 \frac{2\pi}{(-2m)^n} \int_0^\infty\!\! dx\, x^{n+1} I_n(mx) \Gamma_2(x)  \,,
 \ee
where in the integral $x$ is the modulus of $x^\mu$ and $I_n(x)$ is the modified Bessel function of first kind.
We then compute the series expansions
 \begin{equation}
 \widetilde \Gamma_2^{(n)}(-m^2) =m^{2-2n} \left(b_0^{(n)}+ \sum_{k=2}  b_k^{(n)}\,\, g^k \right) , \quad \quad b_0^{(n)}=\delta_{n,1}\,,
\label{eq:Gamma2n-expansion}
 \end{equation}
and plug them in eq.~(\ref{Gamma2m2}) to find the series (\ref{mphExp}).
The values of the coefficients $b_k^{(n)}$ necessary to get the series (\ref{mphExp}) up to ${\cal O}(g^8)$ are reported in table \ref{tab:bkn} in the appendix.
We need to compute $\widetilde \Gamma_2^{(0)}$ up to ${\cal O}(g^8)$, $\widetilde \Gamma_2^{(1)}$ up to ${\cal O}(g^6)$,
$\widetilde \Gamma_2^{(2)}$ up to ${\cal O}(g^4)$ and $\widetilde \Gamma_2^{(3)}$ to ${\cal O}(g^2)$. The order $g^8$ in the two-point function is equivalent
to eight-loops in perturbation theory. The number of topologically distinct non-vanishing graphs for $\Gamma_2(x)$ (in the chosen scheme) as a function of $g$ is reported in table \ref{tab:n-diag-2}.
For instance, up to order $g^3$, we have
\begin{table}[t]
 \centering
 \begin{tabular}{cccccccc}
 \toprule
 $g$ & $g^2$ & $g^3$ & $g^4$ & $g^5$ & $g^6$ & $g^7$ & $g^8$ \\ 
 \midrule
 0 & 1 & 2 & 6 & 19 & 75 & 317 & 1622 \\
 \bottomrule
 \end{tabular}
 \caption{Number of topologically distinct 1PI 2-pt diagrams without self-contractions.}
 \label{tab:n-diag-2}
 \end{table}
 
\bea
\Gamma_2 &=&
 - 96  \vcenter{\hbox{\includegraphics[scale=.4]{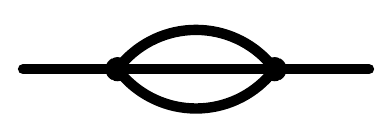}}} \, g^2
 + \left[1152 \vcenter{\hbox{\includegraphics[scale=.35]{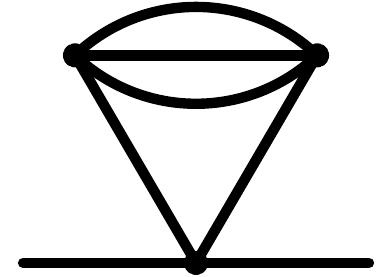}}} 
 + 3456 \vcenter{\hbox{\includegraphics[scale=.4]{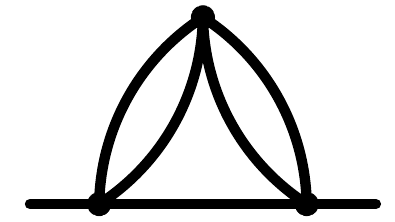}}} 
 \right]  g^3 
 - \left[  41472 \vcenter{\hbox{\includegraphics[scale=.35]{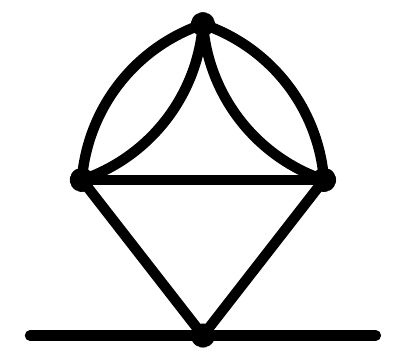}}}
  + 13824 \vcenter{\hbox{\includegraphics[scale=.35]{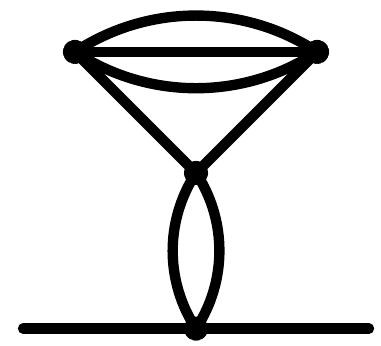}}}
  \right.
  \nn\\
 && ~\left.
  + 82944 \vcenter{\hbox{\includegraphics[scale=.35]{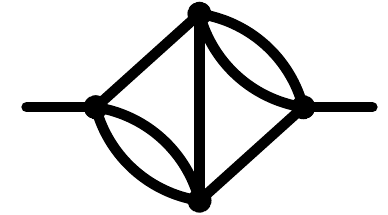}}}
  + 41472 \vcenter{\hbox{\includegraphics[scale=.35]{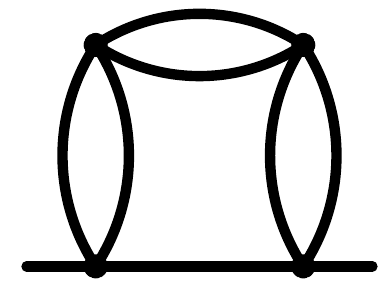}}}
  + 82944 \vcenter{\hbox{\includegraphics[scale=.35]{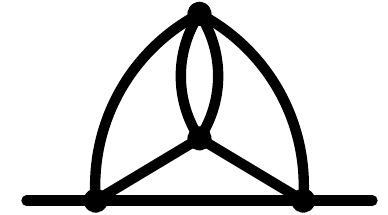}}}
  + 27648 \vcenter{\hbox{\includegraphics[scale=.35]{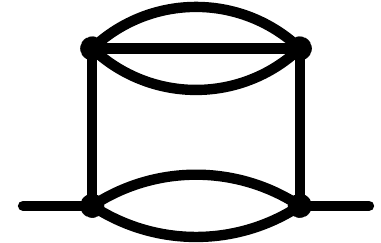}}}
 \right] g^4
 + \, \mathcal{O}(g^5)\,, \nn
\eea

where the numbers indicate the multiplicities of the different topologies.

The final expression for the perturbative expansion of $M^2$ up to $g^8$ order is the following:
\bea
\frac{M^2}{m^2} &= 1 -\frac{3}{2} \, g^2 + \left(\frac{9}{\pi} + \frac{63 \zeta(3)}{2 \pi^3}\right)g^3 - 14.655869(22) \, g^4 
 + 65.97308(43) \, g^5 +\nn \\
 &-347.8881(28) \, g^6 + 2077.703(36) \, g^7 - 13771.04(54) \, g^8  +{\cal O}(g^9) \,.
\label{mphFullSeries}
\eea
As for the vacuum energy, the numbers in parenthesis indicate the statistical errors due to the numerical integration. The coefficients up to order $g^3$ have been computed analytically.
We have also computed the connected two-point function $G_2(x) \equiv \langle \phi(x) \phi(0) \rangle$ in configuration space for various values of $x$.
This requires of course the evaluation of additional Feynman diagrams, connected but not 1PI. 
The knowledge of $G_2(x)$ as a function of $x$ allows us to directly compute, in the critical regime, the exponent $\eta$, see subsection \ref{subsec:criticalreg}.
At fixed order $g^n$, the numerical computation of connected Schwinger functions is more demanding than the  1PI ones, because more integrals have to be performed, due to the presence of the external lines.
For this reason, we have computed $G_2(x)$ up to order $g^6$. We report in table \ref{tab:G2x} in the appendix the coefficients of the series expansion for some selected values of $x$.

\subsection{Large Order Behavior}

It is well known that the large order behavior of the perturbative expansion of $n$-point Schwinger functions $G_n$ in certain QFTs, including 2d and 3d $\phi^4$ theories, can be determined 
by looking at the semi-classical complex instanton configurations \cite{Lipatov:1976ny,Brezin:1976vw,Brezin:1976wa,Brezin:1977gk}. 
In particular, ref.\cite{Brezin:1992sq} worked out the details for the $d=2$ and $d=3$ $O(N)$ vector models (see also ref.\cite{Malatesta:2017xhp} for a recent analysis of next-to-leading large order behavior in the 2d and 3d $\phi^4$ theories). The coupling expansion for the Fourier transform of $n$-point functions at zero momentum $G_n = \sum_k G_n^{(k)} g^k$  behaves, for $k\gg 1$, as
\be
G_n^{(k)}  =  c_n (-a)^k \Gamma(k+b_n+1)\Big(1+{\cal O}(k^{-1}) \Big) \,.
\label{LOB}
\ee
In eq.~(\ref{LOB}), $a$ is an $n$-independent constant, proportional to the inverse of the classical action evaluated at the complex leading instanton configuration,
while $b_n$ and $c_n$ are $n$-dependent parameters that require a detailed analysis of small fluctuations around the instanton configuration.
The coefficient $a$ can only be determined numerically, since the leading instanton solution is not known analytically. For $N=1$, the case of interest, we find
\be
a = 0.683708... \,,
\ee
in agreement with the results of ref.~\cite{Brezin:1992sq}. 
The parameter $b_n$ can be determined analytically and for the 2d $\phi^4$ theory ref.\cite{Brezin:1992sq} finds
\be
b_n= \frac{n}2+1\,.
\label{bLeRoy}
\ee
Ref.\cite{Brezin:1992sq} also determined the coefficient $c_n$, but since this expression is scheme-dependent, we will consider ratios of coefficients so that the dependence on this parameter will cancel out.

The parameters $a$ and $b_n$ play an important role when Borel resumming the perturbative series.
From eq.~(\ref{LOB}) we see that the leading singularity of the Borel transform of $G_n$ sits, for any $n$, at
\be
t = -\frac{1}{a}\,.
\ee
The knowledge of the position of this singularity allows us to use a powerful resummation method known as conformal mapping \cite{loeffel,LeGuillou:1979ixc}, 
that will be reviewed in the next section. 

The large-order estimate (\ref{LOB}) allows us to see how much our truncated series for $\Lambda$ and $M^2$ differ from their asymptotic
behavior. The closer the series is to its asymptotics, the better the resummation methods are to reconstruct non-perturbative results.
To this aim let us define the ratios
\be
r_{n,{\rm asym}}^{(k)} = \frac{G_n^{(k)} }{G_n^{(k-1)} }\,, \quad \quad r_\Lambda^{(k)} = \frac{\Lambda^{(k)}}{\Lambda^{(k-1)}}\,, \quad \quad  r_M^{(k)} = \frac{M^{2(k)}}{M^{2(k-1)}}\,,
\label{ratios}
\ee
where $\Lambda^{(k)}$ and $M^{2(k)}$ are the ${\cal O}(g^k)$ coefficients of the series expansions (\ref{LambdaFullSeries}) and (\ref{mphFullSeries}), and the ratio of ratios
\be
R_\Lambda \equiv \frac{r_{0,{\rm asym}}^{(k)}}{r_\Lambda^{(k)}}\,, \quad \quad R_M \equiv \frac{r_{2,{\rm asym}}^{(k)}}{r_M^{(k)}}\,.
\ee
We report in table \ref{AsymCoeff} $R_\Lambda$ and $R_M$ for different values of $k$.
\begin{table}[t]
\centering
\begin{tabular}{ccccccc} 
\toprule
Order & 3 & 4 & 5 & 6 & 7 & 8 \\
\midrule
$R_\Lambda $ & 2.5059 & 0.9808 & 1.0051 & 0.9941 & 0.9931 & 0.9946 \\
$R_M$ & 1.0040 & 0.9531 & 0.9113 & 0.9076 & 0.9158 & 0.9284\\ 
\bottomrule
\end{tabular}
\caption{Comparison between the ratio of ratios of the series for $\Lambda$ and $M^2$  with their asymptotic values as given by eq.~(\ref{LOB}). 
\label{AsymCoeff}}
\end{table}
It is remarkable how close are the two series to their asymptotic estimates, already for low values of $k$. 
These results should be contrasted to the claimed poor convergence to the asymptotics in the $\epsilon$-expansion, see e.g.\ table VI of ref.\cite{Kompaniets:2017yct}.\footnote{We notice, however, that
ref.~\cite{Kompaniets:2017yct} compares individual coefficients and not their ratios.}

\section{Numerical Determination of the Borel Function and Error Estimate}

\label{sec:numerics}

We have shown in section \ref{BorelGeom} that asymptotic perturbative expansions of Schwinger functions in the 2d $\phi^4$ theory are Borel resummable to the exact result.
In order to be able to concretely use this result, we have to determine the Borel transform ${\cal B}(t)$ of the observable of interest.
The exact form of ${\cal B}(t)$ would require resumming the whole perturbative series, which is clearly
out of reach.  If we naively approximate ${\cal B}(t)$ with its truncated version up to some order $N$ and take the inverse transform term-wise we just get back the original
truncated asymptotic expansion. The root of the problem is easy to understand.
The large-order behavior (\ref{LOB}) indicates that the radius of convergence of the series expansion ${\cal B}(t)=\sum_n B_n t^n$ is $R=1/|a|$. On the other hand, ${\cal B}(t)$ has to be integrated
over the whole positive real axis, beyond the radius of convergence of the series. Obviously
\be
\int_0^\infty \!dt \, e^{-t/g} \sum_{n=0}^\infty B_n t^n \neq \sum_{n=0}^\infty  B_n \int_0^\infty \!dt \, e^{-t/g} t^n  \,.
\label{ineq}
\ee
Any finite truncation of the series, for which sum and integration commute, would result in the right-hand side of eq.~(\ref{ineq}), i.e.\  back to the original asymptotic expansion we started from.\footnote{Notice that 
the integral of the $t^n$ term in eq.~(\ref{ineq}) is dominated by values of $t_0\approx n g$. As long as $n g\lesssim 1/|a|$, the truncated series reliably computes the first terms in a $g$-expansion.
For $n g\gtrsim 1/|a|$, the integral is dominated by values of $t$ beyond the radius of convergence of the series, where its asymptotic nature and the fallacy of the expansion become manifest.}

There are two possible ways to overcome this difficulty: i) manipulating the series to enlarge the radius of convergence over the whole domain of integration or
ii) finding a suitable ansatz for ${\cal B}(t)$ by matching its expansion with the known perturbative coefficients.
In both cases, by knowing a sufficient number of perturbative orders in the expansion, one can approximate
the exact result. We will adopt in the following the conformal mapping and Pad\'e-Borel approximants methods which are arguably 
the most used resummation techniques for cases i) and ii), respectively,\footnote{We call it Pad\'e-Borel method to emphasize that the
approximant is applied to the Borel transform of a function, and not to the function itself.} used to study the 2d $\phi^4$ theory since refs.\cite{Baker:1976ff,Baker:1977hp,loeffel,LeGuillou:1979ixc}.  
We report in what follows some details of these widely used approximation methods to allow the interested reader to reproduce our results.\footnote{Most of the results based on these resummation techniques   in the literature are not very detailed, making hard to reproduce the results  (see ref.\cite{Kompaniets:2017yct} for a recent notable exception in the context of the $\epsilon$-expansion).}  

In both approaches, denoting by $F(g)$ the physical observable of interest, admitting an asymptotic series of the form
\be
\sum_{n=0}^\infty F_n g^n\,,
\label{Fngn}
\ee
we define a generalized Borel-Le~Roy transform as
\be
{\cal B}_b(t) = \sum_{n=0}^\infty \frac{F_n}{\Gamma(n+b+1)} t^n\,,
\label{LeRoy}
\ee
where $b>-1$ is a real parameter. The Borel resummed function is then given by
\be
F_B(g) = \int_0^\infty dt \, t^b e^{-t} {\cal B}_b(g t)\,.
\label{Fg}
\ee
The exact function $F_B(g)$ is independent of $b$, but a dependence will remain in approximations based on truncated series.
As we will see, such dependence can be used both to improve the accuracy of the numerical approximation and to estimate its error.

\subsection{Conformal Mapping}

\label{subsec:CM}

The conformal mapping method  \cite{loeffel,LeGuillou:1979ixc} is a resummation technique that uses in a key way the knowledge of the large order behavior (\ref{LOB}).
After rescaling $t\rightarrow t/g$ so that
\begin{equation}
F_B(g) =  \frac1{g^{1+b}} \int_0^\infty \!\!dt \, t^b e^{-t/g} {\cal B}_b(t) \,,
\end{equation}
the mapping is a clever change of variables of the form\footnote{The mapping (\ref{uTot}) and its inverse (\ref{tTou})  are often seen as mappings of the coupling constant $g$ into some other coupling $w(g)$. Since the argument of the Borel function is the product $tg$, one can equivalently redefine $g$ or $t$. We prefer the 
second choice because it makes more manifest the mapping as a change of variables in the integral.}
 \be
t = \frac 4a \frac{u}{(1-u)^2}\,,
\label{tTou}
\ee
with inverse
\be
u = \frac{\sqrt{1+ a t} -1}{\sqrt{1+ a t} +1}\,.
\label{uTot}
\ee
When we analytically continue ${\cal B}(t)$ in the complex plane, the transformation (\ref{tTou}) turns into a conformal mapping of the plane into a disk of unit radius $|u|=1$.
In particular, under the mapping, the point at infinity in $t$ and the singularity at $t=-1/a$ are mapped at $u=1$ and $u=-1$, respectively.  The branch-cut singularity 
$t\in [-1/a,-\infty]$ is mapped to the boundary of the disc $|u|=1$. Any other point in the $t$-complex plane is mapped within the $u$-unit disc.
The integral $t\in [0,\infty]$ turns into an integral in $u\in[0,1]$. While the series expansion ${\cal B}_b(t)$ has  radius $R=1/|a|$ in $t$, assuming the absence of singularities away from the real axis (we will return to this assumption at the end of this subsection), the series expansion of $\widetilde {\cal B}_b(u)\equiv {\cal B}_b(t(u))$ has  radius $R=1$ in $u$, namely it is convergent over the whole domain of integration. Setting for simplicity the Le~Roy parameter $b=0$, we can now rewrite 
\bea
F_B(g) &= & \frac1g \int_0^\infty \!\!dt \, e^{-t/g} \sum_{n=0}^\infty B_n t^n  =  \frac1g \int_0^1 \!\!du \, \frac{dt}{du} e^{-t(u)/g} \sum_{n=0}^\infty B_n t^n(u)\nn   \\
&=& \frac1g \int_0^1\! \!du \,  \sum_{n=0}^\infty  \widetilde B_n \frac{dt}{du} e^{-t(u)/g}
 u^n 
 =  \frac1g \sum_{n=0}^\infty \widetilde B_n  \int_0^1 \!du \, \frac{dt}{du} e^{-t(u)/g}u^n 
\,,\label{CMtTou}
\eea
where the exchange of the sum with the integral is now allowed. 
Indeed at large $n$ the asymptotic behavior (\ref{LOB}) makes  the coefficients $\widetilde B_n$  polynomially bounded, 
the series in the next-to-last expression of eq.~(\ref{CMtTou}) has coefficients
that are exponentially bounded (as $e^{-3n^{2/3}/(ag)^{1/3}}$) uniformly in $u$, it is therefore possible to find  an integrable function (e.g. a constant) that bounds the series and allows the application of the dominated convergence theorem.
The mapping has converted the asymptotic but divergent series in $g$ into a convergent one!  As can be seen from the form of eq.~(\ref{tTou}), the first $N+1$ terms in $\widetilde B_n$ are easily computed
as linear combinations of the known $N+1$ terms in the original series $B_n$.

In order to further improve the convergence of the $u$-series and at the same time to have more control on the accuracy of the results, we introduce, in addition to the Le Roy parameter $b$, 
another summation variable $s$ \cite{Kazakov:1978ey} and write
\bea
F_B(g) & = & \frac 1g \int_0^\infty \!dt \, \Big(\frac{t}{g}\Big)^b e^{-t/g} \sum_{n=0}^\infty \frac{F_n}{\Gamma(n+b+1)} t^n =
\frac{1}{g^{b+1}}\int_0^1 \!du \, \frac{dt}{du} e^{-t(u)/g} \frac{t^b(u)}{(1-u)^{2s}} \sum_{n=0}^\infty \widetilde B_n^{(b,s)} u^n \nn \\
& =  & \frac{1}{g^{b+1}}\sum_{n=0}^N \widetilde B_n^{(b,s)}  \int_0^1 \!du \, \frac{dt}{du} e^{-t(u)/g} \frac{t^b(u) u^n}{(1-u)^{2s}} +R^{(N)}(g)\equiv F_B^{(N)}(g)+R^{(N)}(g) \,,
\label{FBbs}
\eea
where $R^{(N)}$ is the error associated to the series truncation and $\widetilde B_n^{(b,s)}$ are the coefficients of the Taylor expansion of the function $\widetilde {\cal B}_{b,s}(u)\equiv (1-u)^{2s}\widetilde{\cal B}_b(u)$. The convergence of the series above ensures that $\lim_{N\rightarrow \infty}R^{(N)}=0$.
The use of the modified function $\widetilde {\cal B}_{b,s}$ allows us to better parametrize Borel functions with an arbitrary polynomial behavior at infinity.
At fixed order $N$, for $g\rightarrow \infty$ the integral in eq.~(\ref{FBbs}) is dominated by the region $u\simeq1$. It is immediate to see that in this limit each term in the series behaves similarly, and we have $F_B(g)\sim g^s$ for $g\rightarrow \infty$. Although the exact observable $F(g)$ is independent of the resummation parameters $b$ and $s$, 
a proper choice of $s$, the one that will more closely resemble the actual behavior of $F(g)$ at large $g$, can improve the accuracy of the truncated series.

\begin{figure}[ht]
\centering
\includegraphics[width=0.95\textwidth]{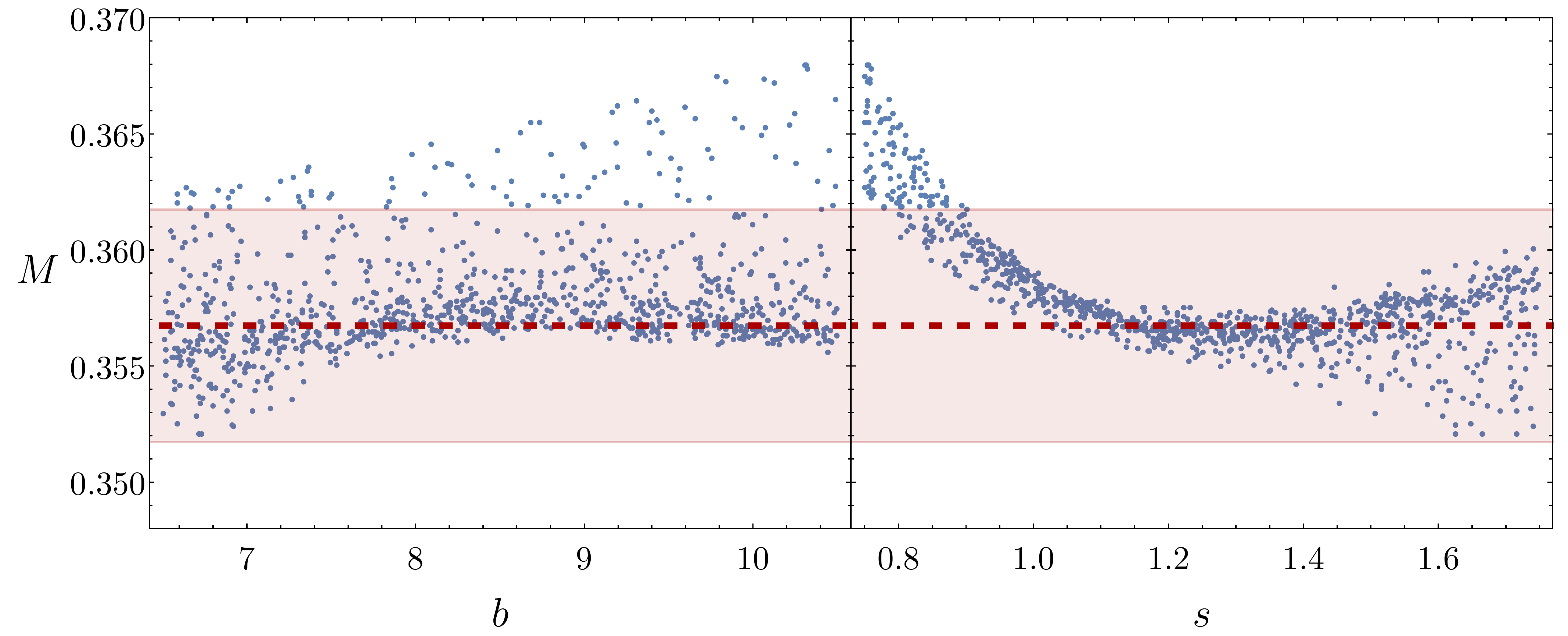}
\caption{Estimate of the error in the resummation of the physical mass at coupling $g=2$. The blue points are the values of $F_{B,k}^{(N=8)}$ as function of the parameters $s$ and $b$. The red dashed line is the central value of the resummation obtained for $s_0=5/4$, $b_0=17/2$. The red band is the final error on the resummation computed as in~\eqref{eq:Err_CM}. 
}
\label{fig:CM-bs-distrib}
\end{figure}

The values of the parameters $b$ and $s$ to use in the resummation are determined by maximizing the convergence of the series and minimizing the sensitivities
to $b$ and $s$. More specifically, we identify a sensible region in the $(b,s)$ parameter space and choose the values $b_0$ and $s_0$ that minimize the following quantity:
\be
\Delta F_B^{(N)} = \sum_{i=s,b} (\partial_i F_B^{(N)})^2+ \Big(|F_B^{(N)}-F_B^{(N-1)}|-|F_B^{(N-1)}-F_B^{(N-2)}|\Big)^2\,.
\label{parafound}
\ee
The first term in eq.~(\ref{parafound}) measures the sensitivity of the observables to $b$ and $s$, while the second term measures how fast the series is converging. 
In order not to disfavor an oscillatory convergence of $F_B^{(N)}$ to a monotonic one, we compute the difference of the differences rather than the simple differences among different loop orders.
Eq.~(\ref{parafound}) is computed at order $N$ for different values of $g$. 
For both the mass and the vacuum energy the values of $s_0^{(N)}$ and $b_0^{(N)}$ do not sensitively depend on the value of $g$ ($s_0$ shows also a mild dependence on $N$, as expected, given its  interpretation in terms of the large coupling behavior of the observable) so we can fix them once and for all for each observable considered.
The central value of our final estimate is given by $F_B^{(N)}(s_0^{(N)},b_0^{(N)})$. 
The final error is taken as follows:
\be
{\rm Err}_{\rm CM}\, F_B^{(N)} = \left(\frac{1}{\Delta s}+\frac{1}{\Delta b} \right)\frac{1}{K}\sum_{k=1}^K 
\left|F_{B,k}^{(N)} - F_{B}^{(N)} \right| + |F_B^{(N)}-F_B^{(N-1)}|\,,
\label{eq:Err_CM}
\ee
where the first term in the above equation takes into account both the residual dependence of the resummation on the parameters $(s,b)$ and the uncertainty in the knowledge of the coefficients (which is however subdominant). It is obtained by generating a set of $K=200$ evaluations $F_{B,k}^{(N)}$ in which the parameters $s_k \in [s_0 - \Delta s,s_0 + \Delta s]$ and $b_k \in [b_0 - \Delta b,b_0 + \Delta b]$ are randomly generated with a flat distribution and the  coefficients of the series are varied with a gaussian distribution around their central value
and with a standard deviation equal to the reported error.
The contribution to the final error is then computed as the mean of the differences between $F_{B,k}^{(N)}$ and the central value $F_{B}^{(N)}$ (in absolute value) weighted by the factor ($1/\Delta s + 1/\Delta b)$, thus reproducing a sort of discrete derivative in the parameter space. 
We find this method more robust than a simple estimate based on the derivatives of $F_{B}^{(N)}$ w.r.t.~$s$ and $b$, which tends to underestimate the error on very flat distributions. 
In the following we set $\Delta s=1/2$ and $\Delta b=2$. 
In fig.~\ref{fig:CM-bs-distrib} we report as an example the points $F_{B,k}^{(N=8)}$ for the physical mass at coupling $g=2$ together with a red band representing the error computed by eq.~\eqref{eq:Err_CM}.

We should emphasize that the error estimate above is subject to some arbitrariness and, as such, it should be seen in a statistical sense as roughly giving one standard deviation from the mean.

As mentioned, there is an important assumption underlying the conformal mapping method. 
All other singularities of the Borel function ${\cal B}(t)$ beyond the one at $t=1/|a|$ should be located on the negative real $t$-axis, so that they are all mapped
at the boundary of the $u$-unit disc. Possible singularities away from the negative $t$-axis would be mapped inside the $u$-unit disc, reducing the radius of convergence of the $u$ series and
invalidating eq.~(\ref{CMtTou}) and the analysis that follows. These singularities would arise from classical solutions to the $\phi$ equation of motion with both a real and an imaginary component. 
The absence of such solutions has been proved in the 1d $\phi^4$  theory (the quartic anharmonic oscillator) but not  in the 2d or 3d $\phi^4$ theories. Using the conformal mapping to our truncated series, 
we do not see instabilities related to the possible presence of such singularities. This justifies, a posteriori, the plausibility of the assumption.

\subsection{Pad\'e-Borel Approximants}

\label{subsec:PB}

Given the first $N+1$ terms of a series expansion of a function 
\be
B(g) = \sum_{n=0}^N B_n g^n + {\cal O}(g^{N+1})\,,
\label{pade1}
\ee
its Pad\'e approximation consists in a rational function of order $[m/n]$
\be
B^{[m/n]}(g) = \frac{\sum_{p=0}^m c_p g^p}{1+\sum_{q=1}^n d_q g^q}\,,
\label{pade2}
\ee
with $m+n = N$. The $m+n+1$ coefficients $c_p$ and $d_q$ are determined by expanding eq.~(\ref{pade2}) around $g=0$  and matching the result up to the $g^N$ term with eq.~(\ref{pade1}).\footnote{The Pad\'e approximation method can in principle be used directly to the observable of interest, rather than to its Borel transform. In so doing the results achieved are often less accurate.
There is an intuitive explanation for that: Pad\'e approximants are manifestly analytic at $g=0$, like the Borel  functions ${\cal B}(g)$. In contrast, the point $g=0$ of $F(g)$
is necessarily singular.} 
Plugging eq.~(\ref{pade2}) in eq.~(\ref{Fg}) leads to an approximation of the observable given by
\be
F_b^{[m/n]}(g) = \int_0^\infty dt \, t^b e^{-t} {\cal B}_b^{[m/n]}(g t)\,.
\label{FgPB}
\ee
The exact Borel function ${\cal B}_b(t)$ is expected to generally have a branch-cut singularity at $t = -1/a$, possible other singularities further away from the origin, 
and no singularities on the real positive axis. This behavior should be reproduced by the location of the poles of the $[m/n]$ approximants.
Unfortunately, quite often $[m/n]$ approximants can show spurious unphysical poles. When such poles are located on the positive real axis, or sufficiently close to it, they give rise to 
large numerical instabilities.\footnote{For $m,n\gg 1$ most of the poles and zeros will accumulate at $t\leq -1/a$, mimicking the presence of a branch-cut singularity (see e.g.\ fig. 4 of ref.\cite{Serone:2017nmd}).  Such large values of $m$ and $n$ cannot be obtained in generic QFTs.}
At a given order $[m/n]$, the location of the poles depends on $b$ and dangerous poles might be avoided by a proper choice of this parameter. If there is no sensible choice
of $b$ that remove them all, the approximant $[m/n]$ is disregarded. For the remaining approximants we proceed as follows.
At fixed $b$ and order $N$, we can have at most $N+1$ different estimates given by $[0/N],[1/(N-1)],\ldots, [N/0]$. Each of them has a different behavior as
$g\rightarrow \infty$:  ${\cal B}_b^{[m/n]}(g) \sim g^{m-n}$.
Suppose the exact Borel function approaches a power like behavior of the form ${\cal B}(g)\sim g^s$ as $g\rightarrow \infty$, leading to $F_B(g)\sim g^s$ in the same limit. If we knew $s$, 
it would be clear that the best Pad\'e-Borel approximant would be given by rational functions of order $[(N+[s])/2,(N-[s])/2)]$, with $[s]$ being the integer value closest to $s$ (and with the appropriate even or odd $N$). As $N$ varies, we would expect that the value of ${\cal B}^{[(N+[s])/2,(N-[s])/2)]}(t)$ as $t\rightarrow\infty$ would be fairly stable and asymptote the value of ${\cal B}(t)$ as $t\rightarrow \infty$.
When $s$ is unknown, like in our case,  the Pad\'e-Borel approximant with the asymptotic behavior closest to the would-be correct one can be found by comparing the $t\rightarrow \infty$ limit of Pad\'e-Borel approximants with the same value of $m-n$ as $N$ varies, and take the series that shows the stablest results \cite{Baker:1977hp}. This procedure can be done scanning over different values of the parameter $b$ or for the selected value of $b$, as explained below. In both cases, the procedure gives always the same answer for the optimal value of $m-n$ to take. Once $m-n\equiv s_0$ is fixed, we 
can consider all the Pad\'e-Borel approximants $[(s_0+n)/n]$ for different values of $n$ that are free of dangerous poles. We then take the highest approximant $[(s_0+n_{\rm max})/n_{\rm max}]$  as our best approximation to the Borel function.\footnote{If $s_0+2n_{\rm max}$ equals $N-1$, rather than $N$, we are effectively not using the $g^N$ coefficient of the series. The latter can however be used to compute $[(s_0+n_{\rm max})/(n_{\rm max}+1)]$ or $[(s_0+n_{\rm max}+1)/n_{\rm max}]$ approximants to test the stability of the result.} 
The value of $s_0$ selected as above is always consistent with the one obtained using the conformal mapping, as explained in 
subsection \ref{subsec:CM}, providing a good consistency check.

The possible presence of spurious poles makes inadequate to fix the parameter $b$ by a scanning procedure like in the conformal mapping case. 
It can however be chosen by knowing the large order behavior of the coefficients (\ref{LOB}). If we take the parameter $b$ in eq.~(\ref{FgPB}) equal to the one 
in eq.~(\ref{bLeRoy}), the Gamma function in eq.~(\ref{LeRoy}) would cancel the similar factor in eq.~(\ref{LOB}), so that the leading singularity of the Borel function at $t=1/a$ is expected to be close to a simple pole\footnote{The singularity is not exactly a pole because of the $1/k$ corrections in eq.~(\ref{LOB}).}
and should be more easily reproduced by Pad\'e-Borel functions that 
can only have pole singularities. If such value of $b$ gives rise to an unstable approximant, we move away from this value until a stable approximant is found at $b=b_0$. 
We then take the Pad\'e-Borel approximant  $F^{[(s_0+n_{\rm max})/n_{\rm max}]}_{b_0}\equiv \hat F_{n_{\rm max}}$ as our best choice.

The error estimate (subject again to some arbitrariness)  is taken as follows:
\be
{\rm Err}\, \hat F_{n_{\rm max}}=  |\partial_{\log b} \hat F_{n_{\rm max}}|_{b_0}+ |\hat F_{n_{\rm max}}-\hat F_{n_{\rm max}-1}| + \Delta^{(s_0+2 n_{\rm max})}\,.
\label{errorPB}
\ee
The term $\Delta^{(s_0+2 n_{\rm max})}$ represents the contribution of the error in the knowledge of the perturbative coefficients.
It is determined by generating a set of 100 random coefficients with a gaussian distribution, with mean and standard deviation as given in eqs.(\ref{LambdaFullSeries}) and (\ref{mphFullSeries}).   
We correspondingly get an approximate gaussian distribution for the 100 output values of $\hat F_{n_{\rm max}}$ and identify $\Delta^{(s_0+2 n_{\rm max})}$ as the standard deviation of this output distribution.  The factor $\Delta^{(s_0+2 n_{\rm max})}$ is typically sub-dominant with respect to the other error terms.

 The conformal mapping method generally gives more accurate results than Pad\'e-Borel approximants. Moreover, the presence of spurious poles hinders a systematic use of the latter.
 We have mostly used Pad\'e-Borel approximants as a consistency check of the results found using the conformal mapping.
An exception is given in section \ref{subsec:massresults} where, in order to extract the critical exponent $\nu$, we resum the inverse of the logarithmic derivative of the physical mass.
In that case the conformal mapping gives poor results and Pad\'e-Borel approximants are preferred. 

\section{Results}

\label{sec:results}

We report in this section the results obtained by resumming the perturbative series as explained in section \ref{sec:numerics}.
We have computed the vacuum energy $\Lambda$ and the physical mass $M$ of the elementary field excitation $\phi$ as a function of the coupling constant $g$, the critical value of the coupling $g_c$ where
the theory is expected to have a second-order phase transition, the critical exponents $\nu$ and $\eta$ and the normalization of the 2-point function at $g=g_c$. From now on we set for simplicity $m^2=1$.

\begin{figure}[t!]
\centering
              \includegraphics[width=78mm]{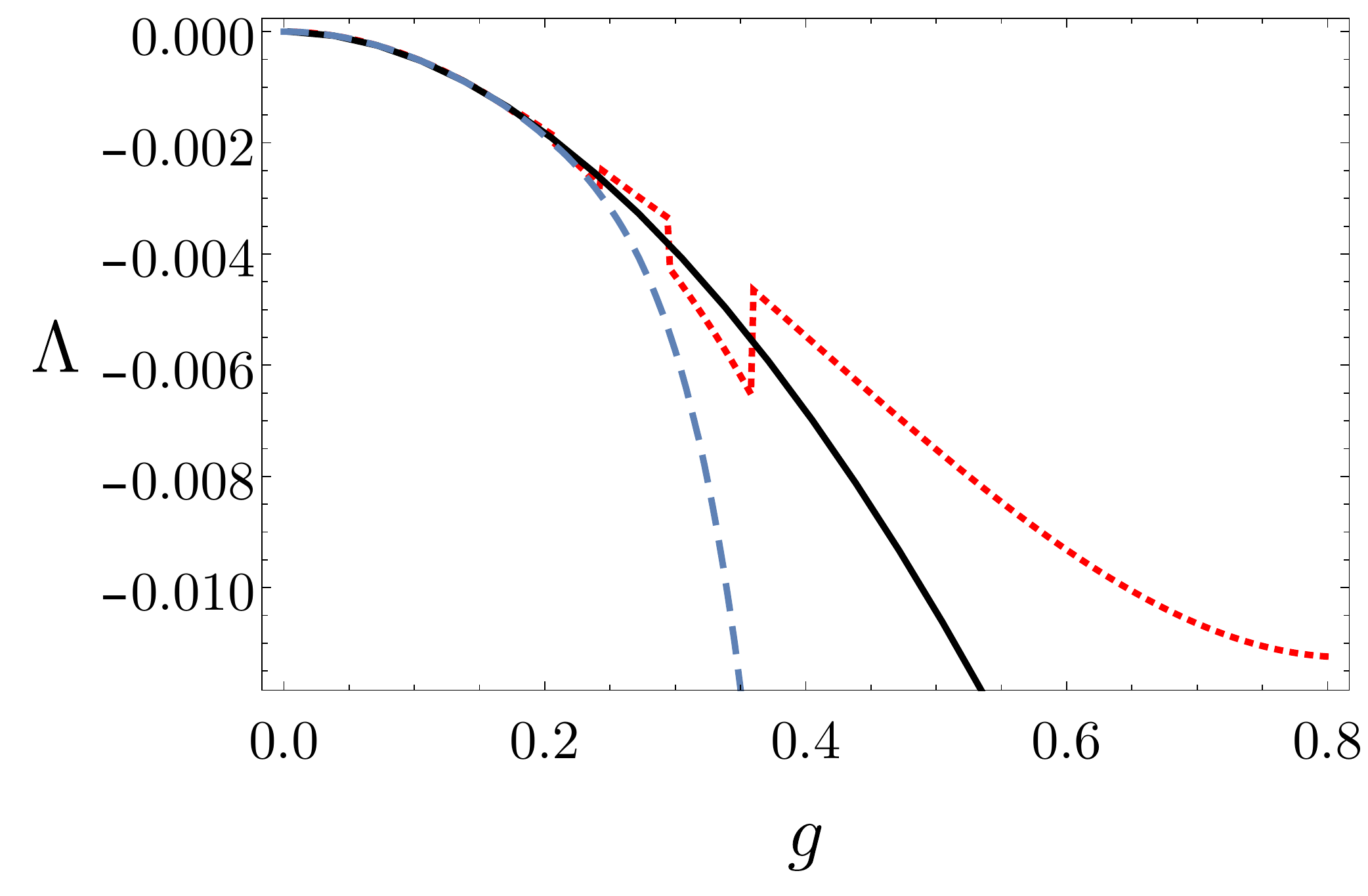}~~
           \includegraphics[width=78mm]{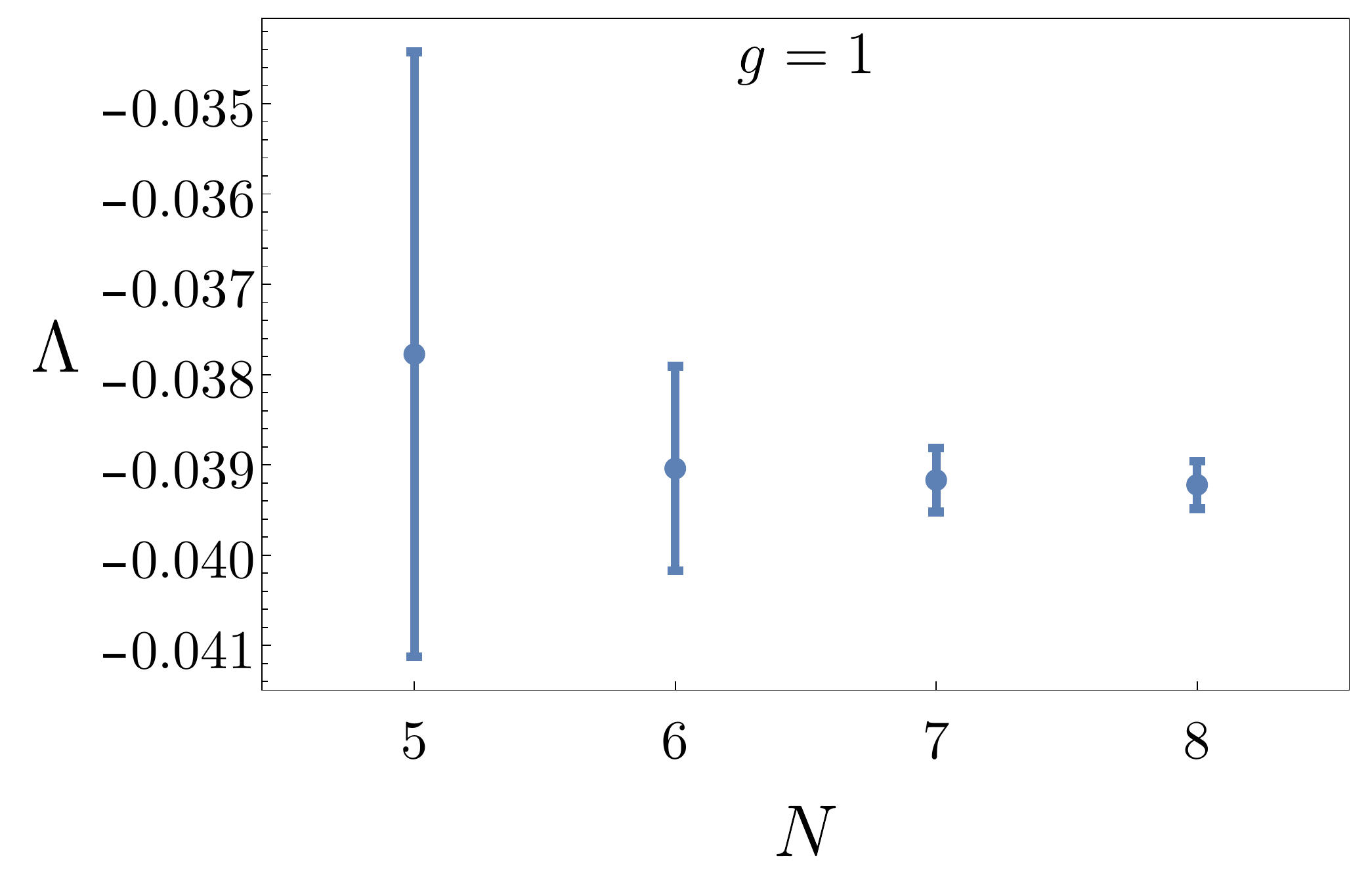}
\caption{\label{fig:LambdaPT}
(Left panel) The vacuum energy $\Lambda$  as a function of the coupling constant $g$ obtained by ordinary perturbation theory up to the $g^8$ order (blue dashed line), optimal truncation (red dotted line) and Borel resummation using conformal mapping (black solid line). Notice how optimal truncation gives accurate predictions up to $g\lesssim 0.4$, in contrast to blind perturbation theory that breaks down for smaller values of the coupling $g\lesssim 0.2$. For $g\gtrsim 0.4$ the optimal truncation estimate breaks down, since we run out of perturbative terms. Errors are not reported to avoid clutter. (Right panel)  Central value and error of $\Lambda(g=1)$ as a function of the $g^N$ terms kept in the conformal mapping resummation technique.}
\end{figure}

\subsection{Vacuum Energy}

In the 2d $\phi^4$ theory, regularized using a normal ordering prescription, the vacuum energy  $\Lambda$  is finite and calculable to all orders in perturbation theory.
 The perturbative expression for $\Lambda$ up to order $g^8$ (i.e.\  nine loops) is reported in eq.~(\ref{LambdaFullSeries}).
We show in the left panel of fig.~\ref{fig:LambdaPT} $\Lambda(g)$ in the weak coupling regime. The asymptotic nature of perturbation theory 
is manifest by comparing ordinary untruncated perturbation theory (blue dashed line) with optimal truncation of perturbation theory (red dotted line).
In the former one keeps all the available perturbative coefficients independently of the value of $g$,  while in the latter the number of terms that are kept in the series expansion changes as $g$ varies, and decreases as $g$ increases.  Indeed, according to the large-order behavior (\ref{LOB}), at fixed $g$ the minimum error in the series expansion is obtained by keeping $N_{\rm Best} \sim 1/(|a| g) \sim 1.5/g$ terms.
For $g\lesssim 0.2$ optimal truncation and untruncated perturbation theory coincides and well approximate the Borel resummed result (black line). For $g\gtrsim 0.2$ untruncated perturbation theory breaks down, while optimal truncation, by removing more and more coefficients in the perturbative expansion, reproduces quite well the Borel resummed result up to $g\approx 0.4$, though with an increasingly unsuppressed error of order $\exp{(-N_{\rm Best})}$.
For $g\gtrsim 0.4$, $N_{\rm Best}$ is so low that optimal truncation becomes unreliable. We conclude that the region $g>0.4$ is inaccessible in perturbation theory.
The value of $g$ where optimal truncation and untruncated perturbation theory differ depends on the order $g^N$ reached and {\it decreases} as $N$ {\it increases}. 
Increasing $N$ by computing more and more terms in the perturbative series would result in a better and better accuracy at small coupling, but
would not change the range of applicability of perturbation theory, which is always $g\lesssim 0.4$, independently of $N$.

In the right panel of fig.~\ref{fig:LambdaPT} we go at strong coupling and report the value of $\Lambda(g=1)$ as a function of the number of coefficient terms used in the conformal mapping method. 
The resummation parameters used are
($s^{(5)}=9/4$, $b^{(5)} =9$),  ($s^{(6)}=5/2$, $b^{(6)} =13/4$),  
($s^{(7)}=5/2$, $b^{(7)} =11/2$), ($s^{(8)}=5/2$, $b^{(8)} = 37/4$). 
The improvement as $N$ increases is evident.

\begin{figure}[t!]
\centering
              \includegraphics[width=78mm]{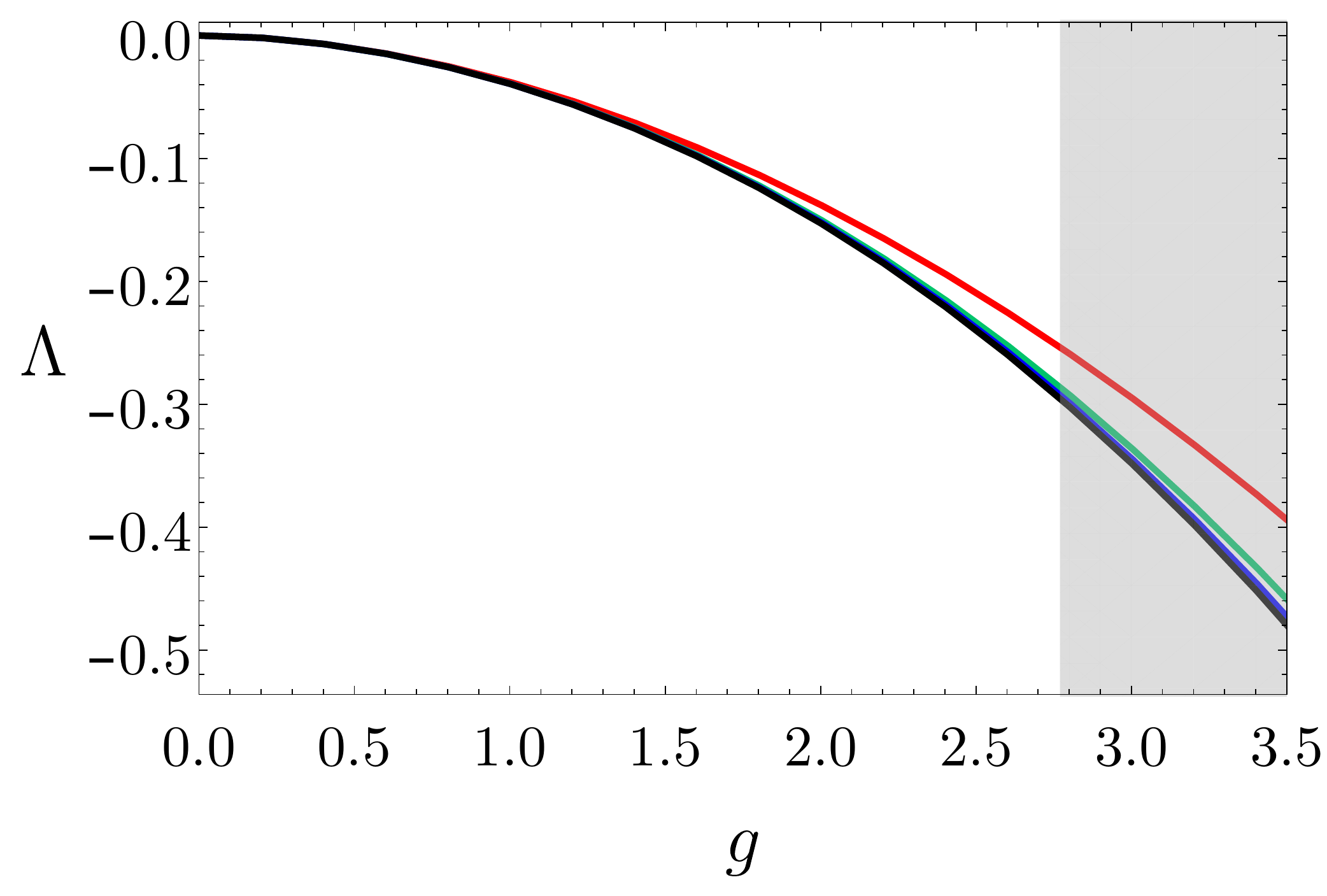}~~
           \includegraphics[width=78mm]{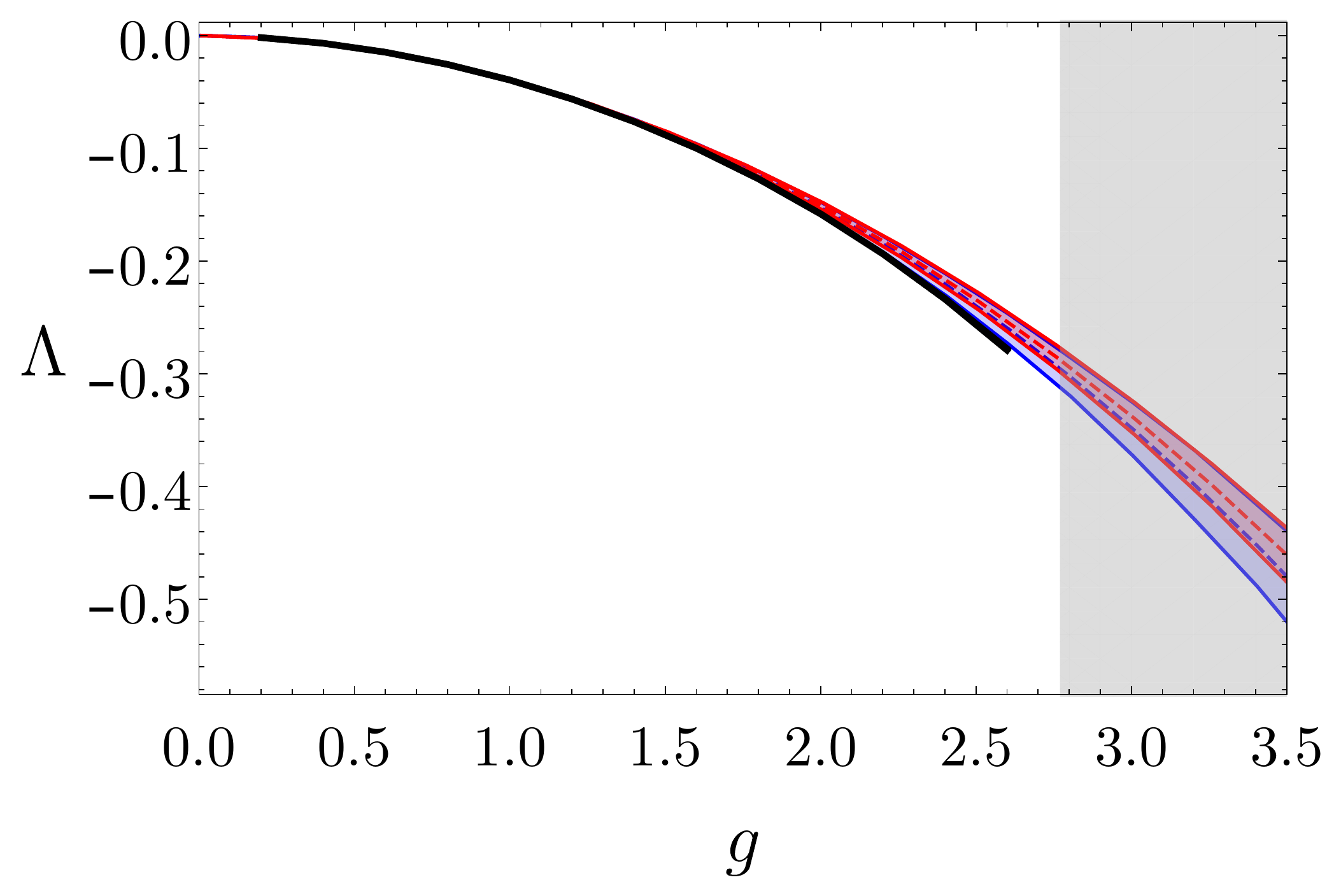}
\caption{\label{fig:LambdaCMLoops}
(Left panel) The vacuum energy $\Lambda$  as a function of the coupling constant $g$ using conformal mapping at different orders: $N=5$ (red line), $N=6$ (green line), $N=7$ (blue line), $N=8$ (black line). Errors are not reported to avoid clutter. The $N=7$ and $N=8$ lines are indistinguishable.
(Right panel) Comparison between the results obtained using conformal mapping at $N=8$ (light blue), Pad\'e-Borel approximants (light red) and the results of ref.\cite{Elias-Miro:2017xxf} (black).}
\end{figure}

In the left panel of fig.~\ref{fig:LambdaCMLoops} we plot $\Lambda(g)$ computed using the conformal mapping at order $N=5,6,7,8$. As $N$ increases, $\Lambda$ tends to bend more towards negative values. The convergence of the result is evident, with the $N=7$ and $N=8$ results overlapping with each other and indistinguishable in the figure. 

In the right panel of fig.~\ref{fig:LambdaCMLoops} we compare the conformal mapping at $N=8$ with the Pad\'e-Borel method and the results of ref.\cite{Elias-Miro:2017xxf}.
In order to avoid dangerous poles, in the Pad\'e-Borel method we have removed the vanishing ${\cal O}(g^0)$ and ${\cal O}(g)$ coefficients from the series and effectively resummed $\Lambda(g)/g^2$.
The approximant shown is $[3/3]$ with $b=-3/4$. All the results are consistent with each other, with the results of ref.\cite{Elias-Miro:2017xxf} consistently at the lower border of our error band for $g\gtrsim 2$, as probably expected, given the $N$-dependence shown in the left panel of fig.~\ref{fig:LambdaCMLoops}.

\subsection{Mass}
\label{subsec:massresults}

\begin{figure}[t!]
\centering
              \includegraphics[width=78mm]{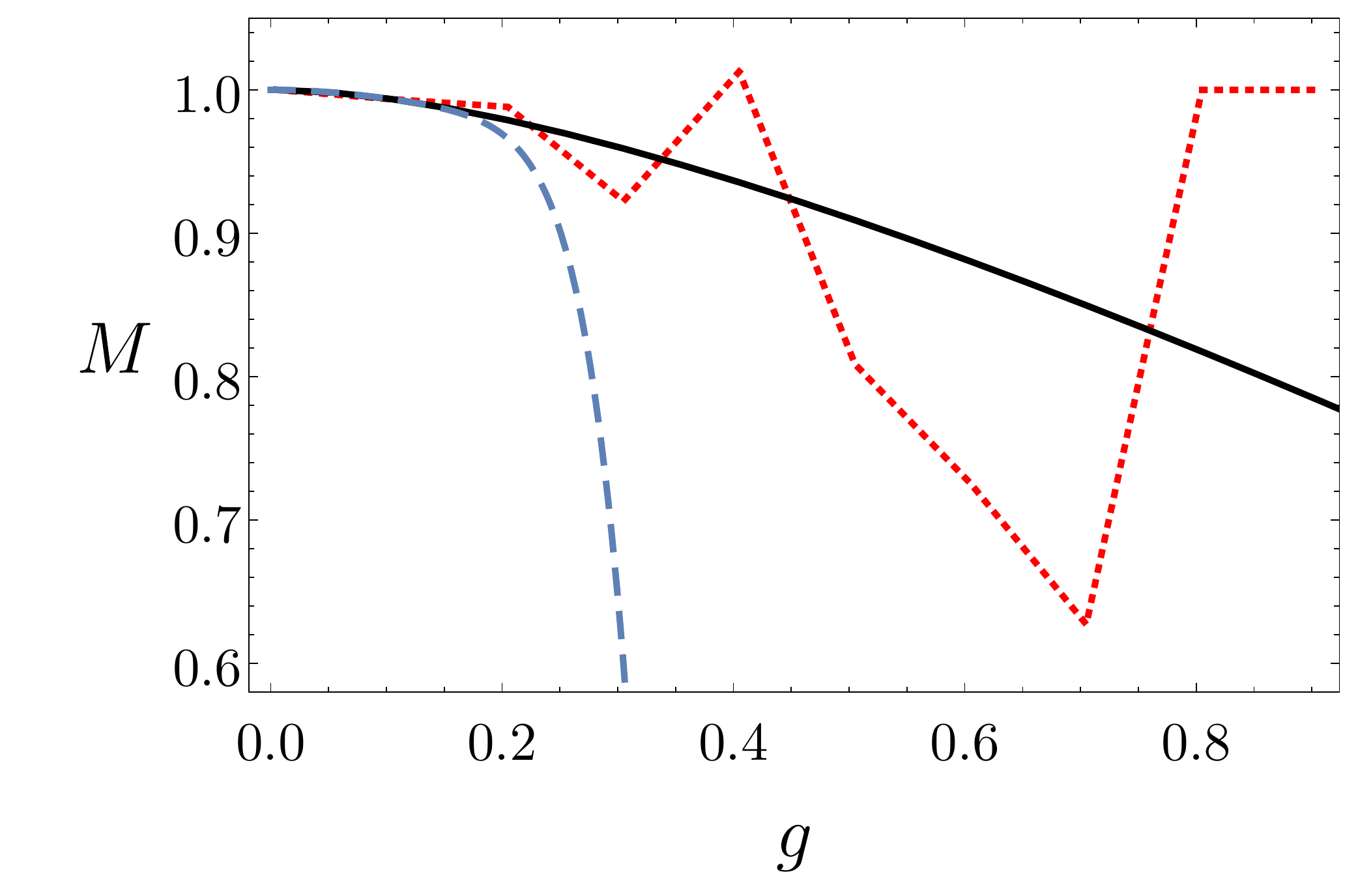}~~
           \includegraphics[width=78mm]{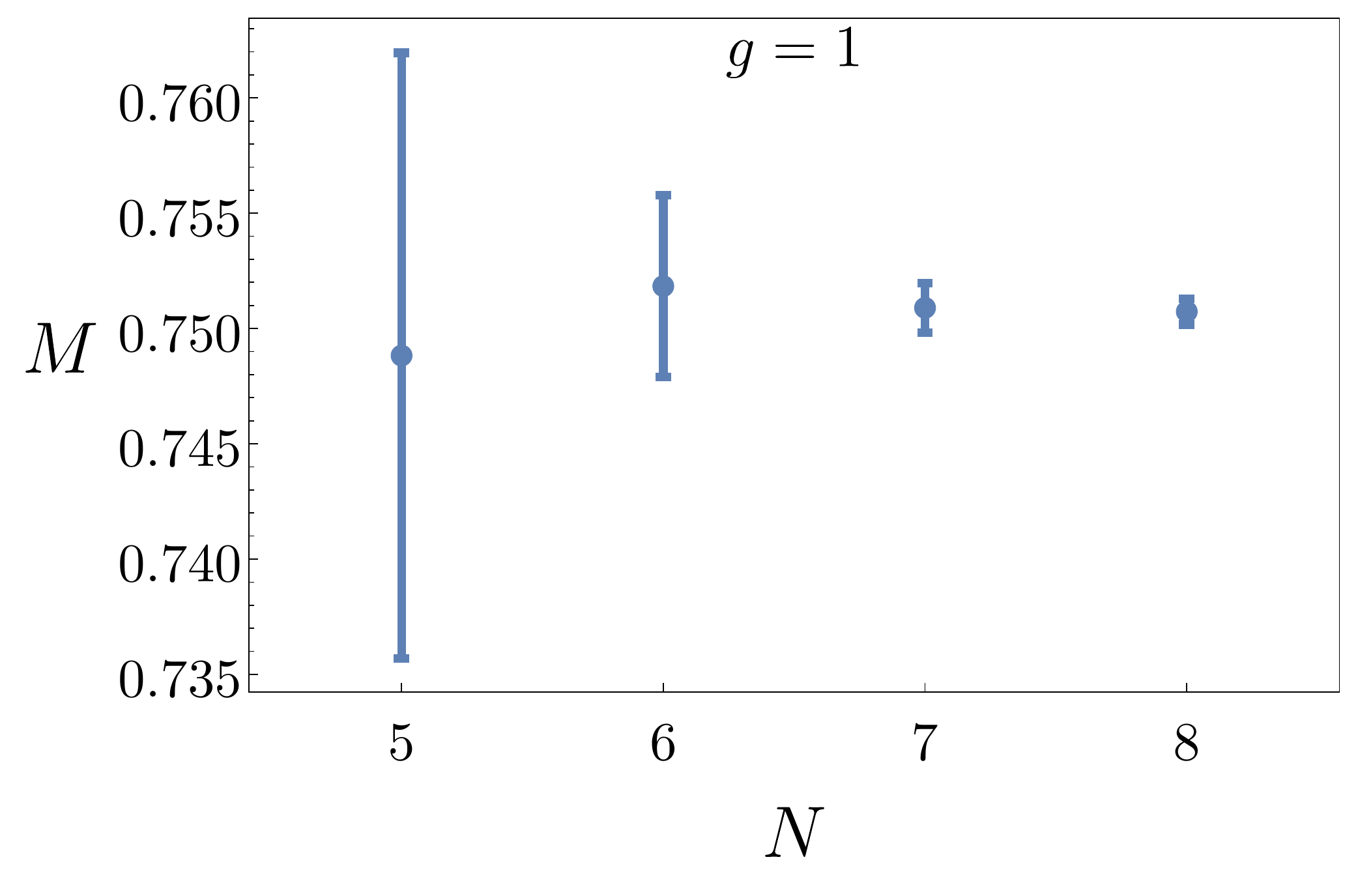}
\caption{\label{fig:MassPT}
(Left panel) The physical mass $M$  as a function of the coupling constant $g$ obtained by ordinary perturbation theory up to the $g^8$ order (blue dashed line), optimal truncation (red dotted line) and Borel resummation using conformal mapping (black solid line). Notice how optimal truncation gives accurate predictions up to $g\lesssim 0.3$, in contrast to blind perturbation theory that breaks down for smaller values of the coupling $g\lesssim 0.15$.  Errors are not reported to avoid clutter. (Right panel)  Central value and error of $M(g=1)$ as a function of the number of loops $N$ kept in the conformal mapping resummation technique.}
\end{figure}

The perturbative expression for $M^2$ up to order $g^8$ (i.e.\  eight loops) is reported in eq.~(\ref{mphFullSeries}).
In the left panel of fig.~\ref{fig:MassPT} we show  $M(g)$ in the weak coupling regime. The asymptotic nature of perturbation theory 
is manifest by comparing ordinary untruncated perturbation theory (blue dashed line) with optimal truncation of perturbation theory (red dotted line).
For $g\lesssim 0.15$ optimal truncation and untruncated perturbation theory coincides and well approximate the Borel resummed result (black line). For $g\gtrsim 0.15$ untruncated perturbation theory breaks down while optimal truncation reproduces quite well the Borel resummed result up to $g\approx 0.3$, again with a quickly increasing error as $N_{\rm Best}$ decreases. For $g\gtrsim 0.3$ optimal truncation becomes unreliable. We conclude that the region $g>0.3$ is inaccessible in perturbation theory.
 
 In the right panel of fig.~\ref{fig:MassPT} we report the value of $M(g=1)$ as a function of the number of coefficient terms used in the conformal mapping method. 
The resummation parameters used are 
$(s^{(5)}=6/5,~b^{(5)} =4)$, $(s^{(6)}=1,~b^{(6)} =4)$,  
$(s^{(7)}=6/5,~b^{(7)} =6)$, $(s^{(8)}=5/4,~b^{(8)} = 17/2)$.  
The improvement as $N$ increases is evident.

\begin{figure}[t!]
\centering
              \includegraphics[width=78mm]{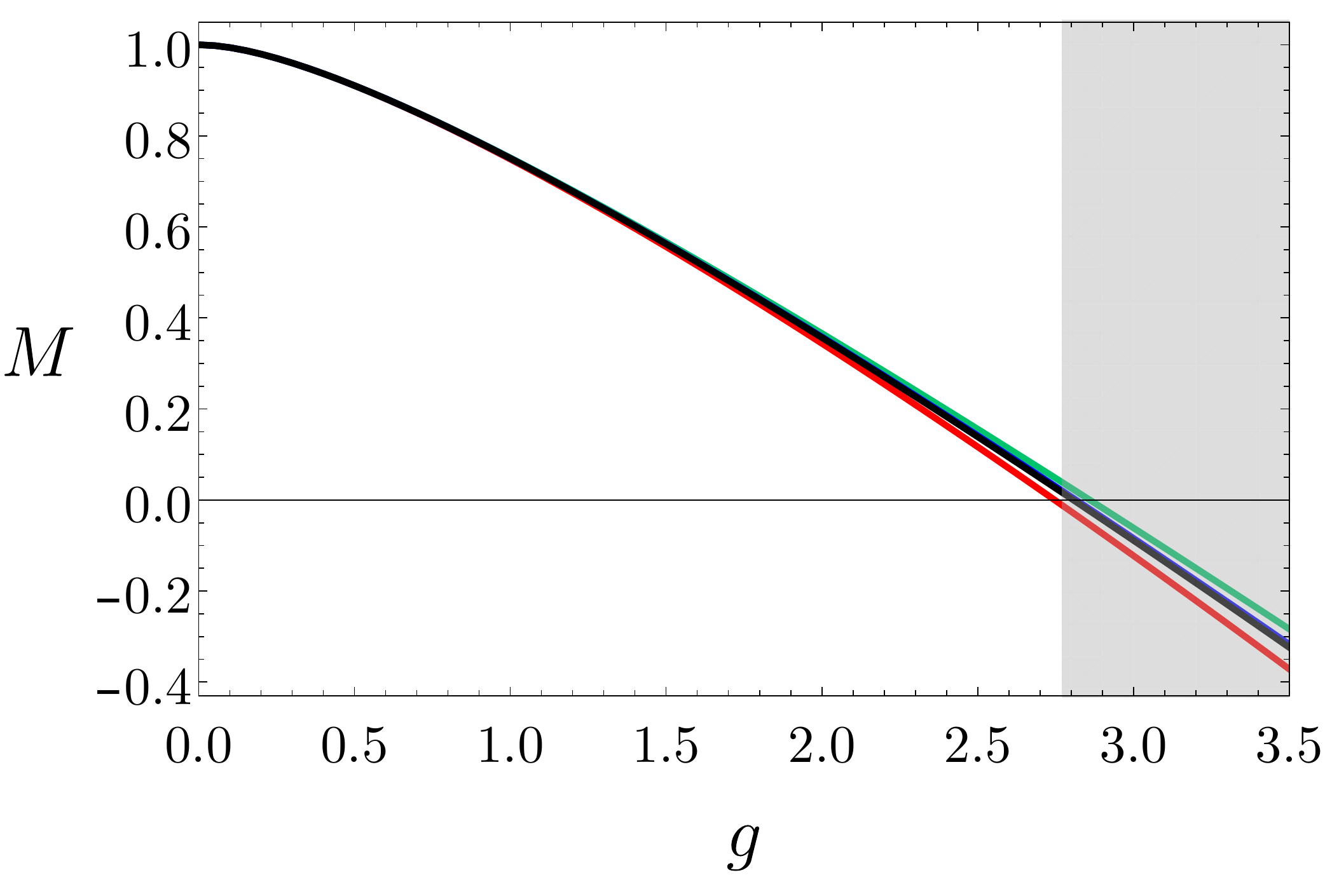}~~%
           \includegraphics[width=78mm]{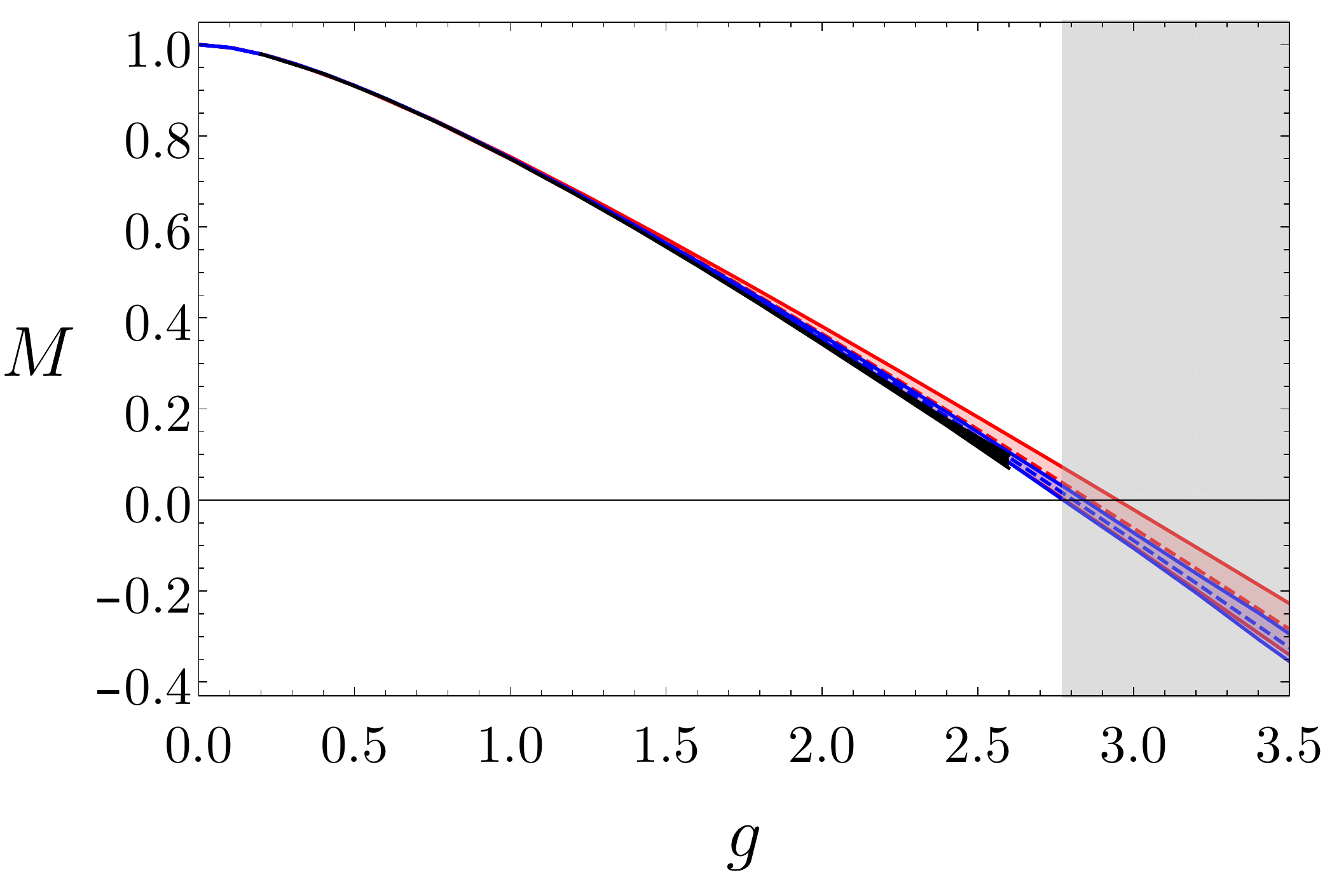}%
\caption{\label{fig:MassCMLoops}
(Left panel) The physical mass $M$  as a function of the coupling constant $g$ using conformal mapping at different orders: $N=5$ (red line), $N=6$ (green line), $N=7$ (blue line), $N=8$ (black line). Errors are not reported to avoid clutter. The $N=7$ and $N=8$ lines are indistinguishable.
(Right panel) Comparison between the results obtained using conformal mapping at $L=8$ (light blue), Pad\'e-Borel approximants (light red) and the results of ref.~\cite{Elias-Miro:2017xxf} (black).}
\end{figure}

In the left panel of fig.~\ref{fig:MassCMLoops} we plot $M(g)$ computed using the conformal mapping method at order $N=5,6,7,8$. The convergence of the result is evident, with the $N=7$ and $N=8$ results overlapping 
and indistinguishable in the figure. 
 
In the right panel of fig.~\ref{fig:MassCMLoops} we compare the conformal mapping at $N=8$ with the Pad\'e-Borel method and the results of ref.~\cite{Elias-Miro:2017xxf}.
The approximant shown is $[4/3]$ with $b=1$. All the results are nicely consistent with each other in the whole range of couplings shown. 

\subsection{Critical Regime}

\label{subsec:criticalreg}

\begin{figure}[t!]
\centering
\includegraphics[height=50mm]{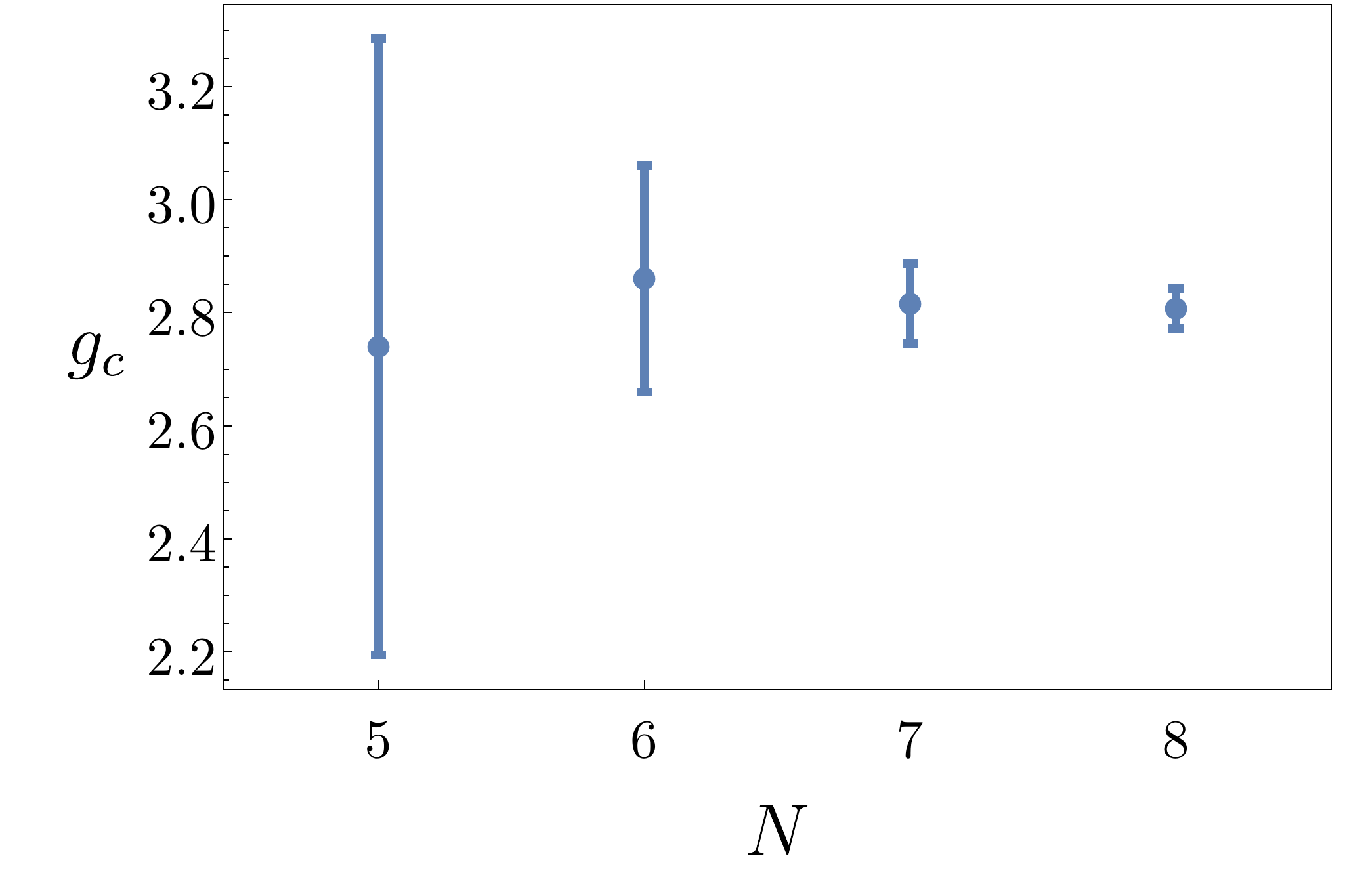}
\hspace{.5cm}
\includegraphics[height=50mm]{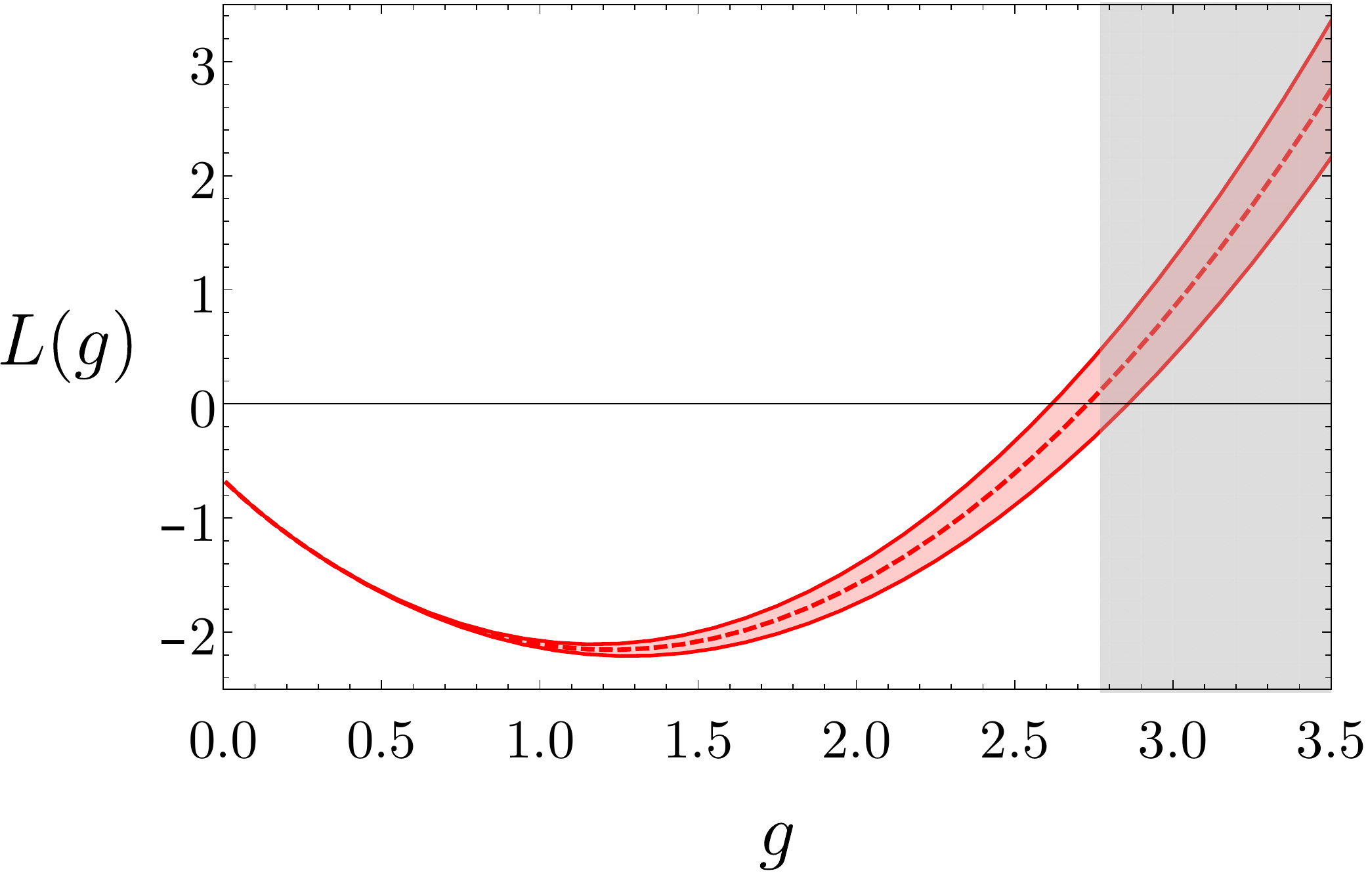}%
\caption{\label{fig:LogMass} 
(Left panel) Central value and error of $g_c$ as a function of the number of loops $N$ kept in the conformal mapping resummation technique.
(Right panel) The inverse of the logarithmic derivative of the physical mass $M$  as a function of the coupling constant $g$ using a $[4/2]$ Pad\'e-Borel approximant. 
This expression allows us to determine the critical exponent $\nu$, as explained in the main text. }
\end{figure}

The critical coupling $g_c$ where the second-order phase transition occurs is determined as 
\be
M(g_c)=0\,.
\ee  
We report in the left panel of fig.~\ref{fig:LogMass} the value of $g_c$  using conformal mapping at different orders. Our final estimate is given by
\be
g_c = 2.807(34)\,.
\label{eq:g_crit}
\ee
We now turn to the determination of the critical exponents $\nu$ and $\eta$.
The exponent $\nu$, defined in eq.~(\ref{nuDef}), is known to be exactly equal to 1 in the 2d Ising model, which is in the same universality class of the critical 2d $\phi^4$ theory.
In some sense, we have tacitly used before the information $\nu=1$ in deciding to resum $M$ instead of, say, $M^2$ or some other function of $M$.
As a matter of fact,  the accuracy of the results significantly depend on which mass function one decides to resum and 
the best choice should be given by the function that approaches the critical point smoothly, and has a simple zero at $g=g_c$, i.e.\  $M(g)$.
This expectation is fully confirmed by our analysis. Since the Borel resummed mass is an analytic function of the coupling, 
by resumming $M(g)$ we would automatically, but somewhat trivially, obtain $\nu=1$.

A possible way to determine $\nu$ (and $g_c$) is obtained by
resumming the combination 
\be
L(g)\equiv \frac{2g^2}{g \partial_g \log M^2}.
\ee 
Close to the critical point, for $g\rightarrow g_c^-$, we have
\be
L(g)= \frac{g_c}{\nu} (g-g_c) + {\cal O}\Big((g_c-g)^2\Big)\,,
\ee
and $\nu$ can be extracted as
\be
\nu = \frac{g_c}{\partial_g L}\bigg|_{g=g_c}\,.
\label{nuextr}
\ee
In fig.~\ref{fig:LogMass} we report $L(g)$ as a function of $g$ obtained by a $[4/2]$ Pad\'e-Borel approximant, the maximal approximant of the form $[m+2/m]$ for $L(g)$ that has an expansion up to $O(g^6)$.
The conformal mapping technique does not give good results for $L(g)$, probably because the coefficients in the series expansion of $L(g)$
differ more from the asymptotic values (\ref{LOB}) due to the manipulation of taking the inverse of a logarithmic derivative, and this results in a poor accuracy.

The values of the critical coupling and $\nu$ obtained from $L(g_c)=0$ and eq.~(\ref{nuextr}) are
\be
g_c = 2.73(12) \,, \quad \quad \nu = 0.96(6)\,.
\ee
The result for $g_c$ is consistent with that obtained in eq.~(\ref{eq:g_crit}) by resumming $M$, but it has a larger uncertainty, so we take as our best estimate the result (\ref{eq:g_crit}).

\begin{figure}[t!]
\begin{center}
              \includegraphics[width=95mm]{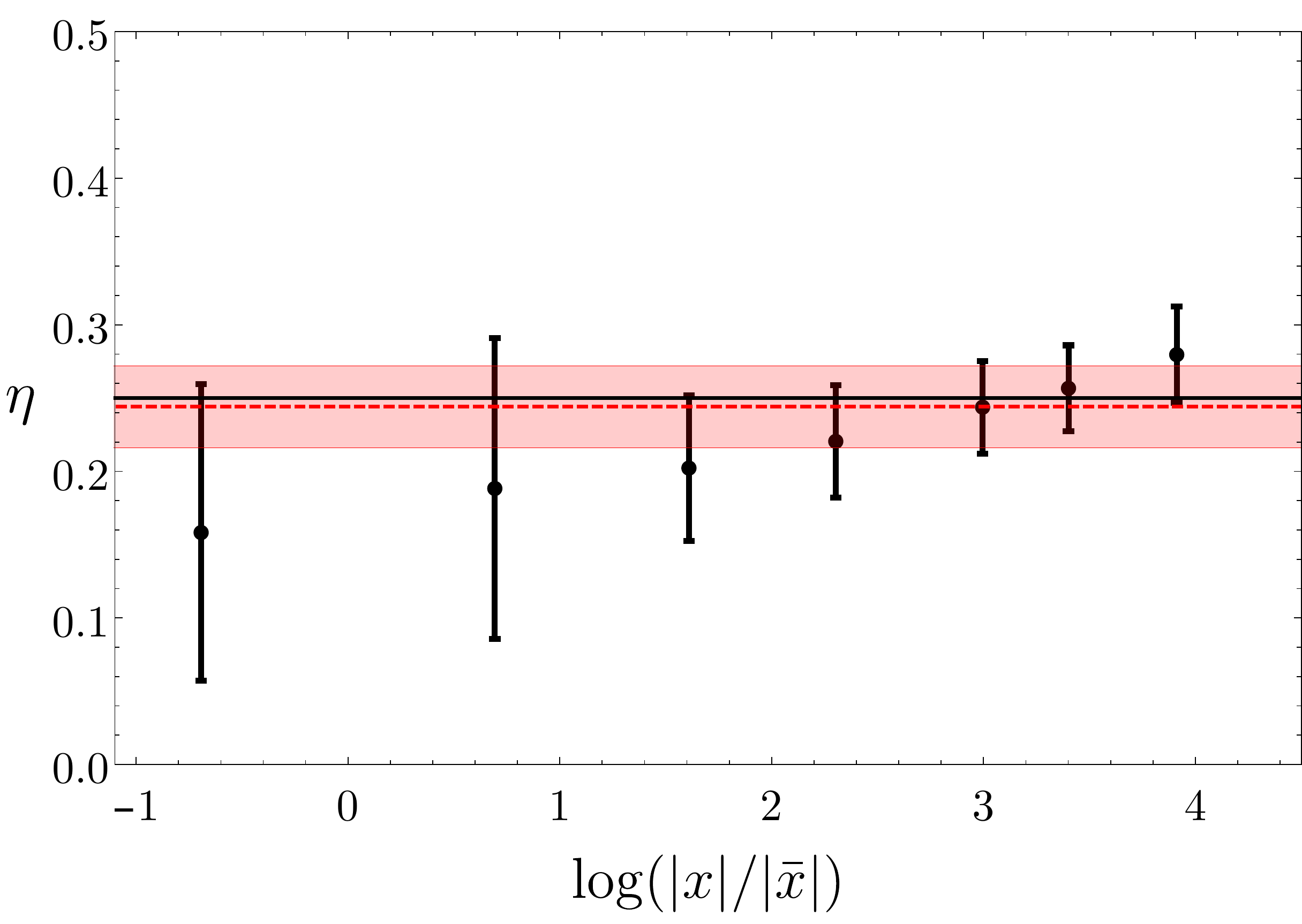}\\
\end{center}
\caption{\label{fig:etafit}
The slope coefficient $-\log[T(x,\bar x)]/\log(|x|/|\bar x|) $ of the normalized 2 point function computed by Borel resumming the series at $g=g_c$, as a function of $\log(|x|/|\bar x|)$.
At the critical point it is supposed to be constant in $x$ and equal to the critical exponent $\eta$. The black solid line represents the theoretically known value $\eta = 1/4$.}
\end{figure}

The exponent $\eta$ is defined in eq.~(\ref{etaDef}).  
As well-known, the field $\phi(x)$ at criticality flows to the magnetization field $\sigma$ of the two-dimensional Ising model and one has $\eta = 2\Delta_\sigma = 1/4$.
We provide a perturbative estimate of $\eta$ as follows: we define a normalized two-point function
\be
T(x,\bar x)\equiv \frac{\langle \phi(x) \phi(0) \rangle}{\langle \phi(\bar x) \phi(0) \rangle}
\ee
where $\bar x$ is an arbitrary fixed value. We then calculate $T(x,\bar x)$ from the Borel resummed two-point function $\langle \phi(x) \phi(0)\rangle $ 
and evaluate it at $g_c$.
In this way, $\eta$ is given by
\be \label{eq:etafrom2pt}
\eta = - \frac{\log T(x,\bar x)}{\log (|x|/|\bar x|)}\,.
\ee
In fig.~\ref{fig:etafit} we plot the results obtained for the ratio of eq.~(\ref{eq:etafrom2pt})
as a function of $\log(|x|/|\bar x|)$ for different values of $x$ at $g=g_c$. 
We use the value of $g_c$ as determined in eq.~\eqref{eq:g_crit} and we fix $|\bar x|=1/100$. 
The points are compatible with a constant giving
\be
\eta = 0.244(28) \,.
\ee
The uncertainty obtained by varying the critical coupling $g_c$ within its error is subleading.
The base-point $\bar x$ has been chosen as the point for which the correlator $\langle \phi(\bar x) \phi(0) \rangle$ at $g=g_c$ has the smallest error, thus minimizing the errors on $T(x,\bar x)$. In practice choosing a different $\bar x$ has only a very small effect on the determination of $\eta$, with the central value estimate always being well within the reported error.
Because of the uncertainty $\Delta g$ in $g_c$, the two-point function would still have an exponential suppression roughly of order $\exp(-M(\Delta g)|x|)$. 
Requiring $M(\Delta g) |x|\ll 1$, so that the exponential factor does not spoil our determination of $\eta$, leads to a bound 
$|x|<1$ in the selection of points.

Once the exponent $\eta$ is known, we can also extract the two-point function normalization $\kappa$ using eq.~(\ref{etaDef}):
\be
\kappa = |x|^\eta \langle \phi(x) \phi(0) \rangle_{g=g_c}  \,
\label{Ztwopoint}
\ee
by taking the mean of the values at different $x$. We obtain $\kappa=0.29(2)$ where the reported error takes into account both the uncertainty in the determination of the exponent $\eta$ and the uncertainty in the resummation.

\section{Comparison with Other Approaches} 

\label{sec:comparisons}
 
The critical value of the coupling $g_c$ in the 2d $\phi^4$ theory has been determined using a variety of approaches, such as lattice Monte Carlo, lattice matrix product states, Hamiltonian truncations
and variants of the resummation of perturbation theory (supplemented by lattice) performed in this work.
Most of these approaches, including our work, are based on an ordinary covariant quantization
and normal ordering regularization, making possible  a direct comparison of a scheme-dependent quantity like $g_c$.
A comparison with approaches using other quantization or regularization schemes, such as the light-cone Hamiltonian truncation methods of refs.\cite{Burkardt:2016ffk,Anand:2017yij},
requires a careful mapping of the parameters and will not be considered (see ref.\cite{Fitzpatrick:2018ttk} for a recent attempt in this direction).

\begin{table}[t]
\centering
\begin{tabular}{c c l l}
\toprule
Ref.  &    &  ~~~$g_c$  &    ~~~Method     \\
\midrule \\[-16.7pt]
{This work} &  \multirow{7}{*}{\includegraphics[scale=.205]{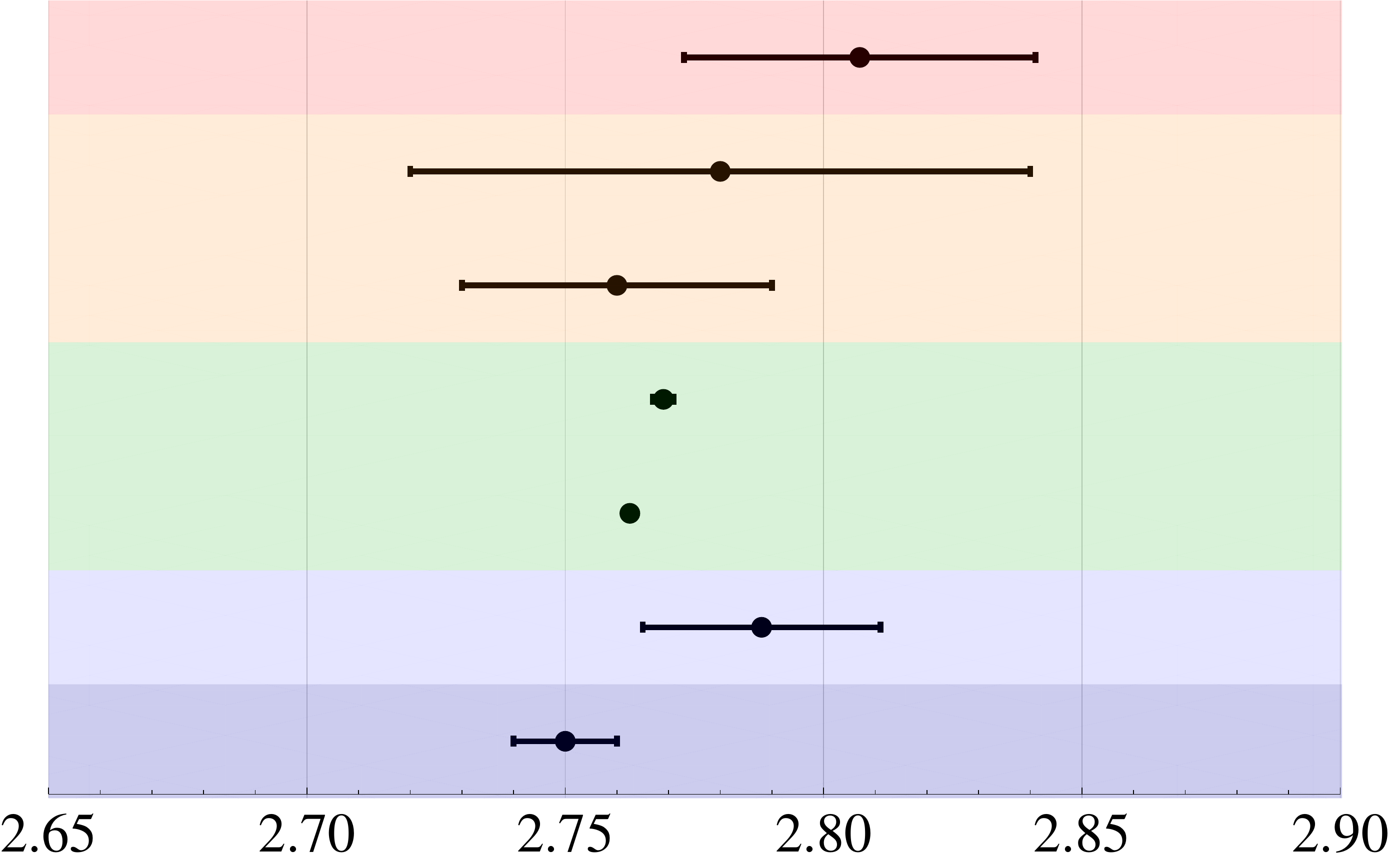}} & $2.807(34)$  &  Borel RPT \\
\cite{Bajnok:2015bgw} &&  $2.78(6)$  & \multirow{2}{*}{HT} \\
\cite{Elias-Miro:2017xxf,Elias-Miro:2017tup} && $2.76(3)$  &  \\
\multirow{2}{*}{\cite{Milsted:2013rxa}} &&  $2.769(2)$   & MPS I \\
 & & $2.7625(8)$  & MPS II \\
\cite{Bosetti:2015lsa} & & $2.788(15)(8)$     & LMC \\
\cite{Pelissetto:2015yha} &&  $2.75(1)$  & LMC+Borel RPT \\[-3pt]
\midrule \\[-8pt] \bottomrule
\end{tabular}
\caption{Computation of $g_c$ using Borel Resummed Perturbation Theory (RPT), Hamiltonian Truncation (HT), Matrix Product States (MPS) and Lattice Monte Carlo (LMC) methods.} 
\label{tablegc}
\end{table}

We report in table \ref{tablegc} the most recent results for $g_c$ using various methods. The Hamiltonian truncation results in refs.\cite{Elias-Miro:2017xxf,Elias-Miro:2017tup} and \cite{Bajnok:2015bgw} are based on the study of the ${\mathbb Z}_2$ unbroken and broken phases, respectively. The lattice analysis in ref.\cite{Milsted:2013rxa} is based on a tensor network with matrix product states.
Ref.\cite{Milsted:2013rxa} reports two values for $g_c$, denoted by I and II in the table: in I $g_c$ is defined as the value where the energy of the first excited state vanishes, while
II is obtained by looking at the value where $\langle \phi \rangle$ vanishes, starting from the ${\mathbb Z}_2$ broken phase.
Ref.\cite{Bosetti:2015lsa} is based on a Monte Carlo simulation. Finally, ref.\cite{Pelissetto:2015yha} uses lattice results to express the critical value of the renormalized coupling $g_R$, obtained by resumming the perturbative series in $g_R$ \cite{Baker:1976ff,Baker:1977hp,Pelissetto:1999cr} (see next section), in terms of the coupling $g$.

In table \ref{tablegLambdaM}  we compare the values of $\Lambda$ and $M$ for three values of the coupling $g$ with those obtained in refs.\cite{Elias-Miro:2017xxf,Elias-Miro:2017tup}.
At weak coupling $g=0.2$, as probably expected, we get more accurate results but as the coupling increases (at $g=1$ and at $g=2$) the accuracy reached by refs.\cite{Elias-Miro:2017xxf,Elias-Miro:2017tup}
is better than ours. Overall, given also the statistical nature of our error, the results are in very good agreement between each other.

\subsection{Comparison with Other Resummation Methods}

\label{subsec:comparisonresum}

As mentioned in the introduction, two different resummation methods have already been developed in the literature: resummation in the $\epsilon$-expansion \cite{Wilson:1971dc} and at fixed dimension \cite{Parisi:1993sp}. The latter differs from our perturbative expansion in the renormalization scheme and the
fact that the critical point is extracted from the vanishing of the $\beta$ function (involving the 4-point function) instead of the mass gap (only involving the 2-point function), 
for this reason we denote such expansion PT$_{\rm 4pt}$, while ours will be called  PT$_{\rm 2pt}$.

\begin{table}[t]
\centering
\begin{tabular}{ccll}
\toprule
Ref. &  $g$  &  \multicolumn{1}{c}{$\Lambda$}  &  \multicolumn{1}{c}{$M$}  \\
\midrule
This work &  $0.2$  &  $-0.00181641(8)$ &  $0.9797313(4)$ \\
\cite{Elias-Miro:2017xxf,Elias-Miro:2017tup} &  $0.2$  & $-0.0018166(5)$  & $0.979733(5)$ \\
\midrule
This work &  $1$ & $-0.0392(3)$ &  $0.7507(5)$ \\
\cite{Elias-Miro:2017xxf,Elias-Miro:2017tup} & $1$  & $-0.03941(2)$ & $0.7494(2)$\\
\midrule
This work & $2$ & $-0.153(5)$ &  $0.357(5)$ \\
\cite{Elias-Miro:2017xxf,Elias-Miro:2017tup} & $2$ & $-0.1581(1)$ & $0.345(2)$ \\
\bottomrule
\end{tabular}
\caption{The values $\Lambda$ and $M$ for different values of $g$ and comparison with refs.\cite{Elias-Miro:2017xxf,Elias-Miro:2017tup}.} 
\label{tablegLambdaM}
\end{table}

The $\epsilon$-expansion is devised to study the critical theory as a function of the space-time dimensions. It is a well-known technique and has been used in a variety of contexts.
Within applications to the 2d $\phi^4$ theory, one first determines the value of the critical coupling $g^*$ as a function of $\epsilon=4-d$ perturbatively from the $\beta$-function of the quartic coupling, 
computes critical exponents by resumming the corresponding series in $g^*(\epsilon)$ and then set $\epsilon=2$ \cite{LeGuillou:1985pg}. 

The fixed dimension coupling expansion of ref.\cite{Parisi:1993sp} (see e.g.\ chapters 26 and 29 of ref.\cite{ZJBook} for an introduction) is based on an expansion in terms of a renormalized dimensionless coupling $g_R$.
As we mentioned, in the 2d  $\phi^4$ theory  there is no need to introduce a wave function renormalization constant $Z$ for the field $\phi$ or a coupling counterterm, since the bare coupling constant $\lambda$ is finite. Nevertheless, one can define renormalized quantities in analogy to the 4d $\phi^4$ renormalization conditions. Denoting by $\Gamma_R^{(n)}$ 
the 1-particle irreducible (1PI) $n$-point renormalized Schwinger functions, one defines $m_R$, $Z$ and $g_R$ by the following three conditions at zero momentum:\footnote{Notice the different normalization of the coupling: $g_R = 4! g +{\cal O}(g^2)$.}
\be
\Gamma_R^{(2)}(p=0)= m_R^2\,, \quad \frac{d\Gamma_R^{(2)}(0)}{dp^2} =1\,, \quad \Gamma^{(4)}_R(0) = m_R^2 g_R\,,
\label{RenCond}
\ee
where as usual $\Gamma_R^{(n)}$ are related to the bare 1PI Schwinger functions $\Gamma^{(n)} $ as $\Gamma_R^{(n)}=\Gamma^{(n)} Z^{n/2}$. In the critical regime $m_R \rightarrow 0$ the
$\Gamma_R^{(n)}$ are expected to satisfy an homogeneous Callan-Symanzik equation in terms of a $\beta$-function defined as 
\be
\beta(g_R) = m_R \frac{d g_R}{dm_R}\bigg|_\lambda  \,.
\label{betaphi24}
\ee
In the proximity of a fixed point  $g_R^*$  where $\beta(g_R^*)=0$, the Schwinger functions would satisfy the typical scaling behavior of a critical theory.
In contrast to the $\epsilon$-expansion case, the fixed point cannot be accessed perturbatively and is determined by Borel resumming the truncated expansion of $\beta(g_R)$.
Once $g_R^*$ is determined, critical exponents can be computed like in the $\epsilon$-expansion by Borel resumming their series in $g_R$, setting $g_R=g_R^*$ after the resummation.

Using our perturbative coefficients of the 2-pt and 4-pt functions (see the  Appendix) and the above procedure, the expansions for $\beta(g_R)$ and $\eta(g_R)$ read as follows:\footnote{Using an appropriate Callan-Symanzik equation, the critical exponent $\nu$ could also be extracted from our perturbative coefficients. We thank Riccardo Guida for this observation.}
{\small{\bea
\frac{\beta(v)}{2}  & = &-v + v^2 - 0.7161736 v^3 + 0.930768(3) v^4 - 1.5824(2) v^5 + 3.2591(9) v^6 - 7.711(5) v^7 \nn \\
&& + 20.12(9) v^8 + \mathcal{O}(v^9)%
 \,,\nn\\ 
\eta(v) & = &  0.0339661 v^2 - 0.0020226 v^3 + 0.0113932(2) v^4 - 0.013735(1) v^5 + 0.02823(1) v^6   \nn \\
&& - 0.06179(9) v^7 + 0.1475(8) v^8 + \mathcal{O}(v^9)\,, \label{betaeta}
\eea}}
where we have defined $v=3g_R/(8\pi)$. The first 5(6) coefficients in $\beta(v)$ and the first 3(4) in $\eta(v)$ agree with the results obtained in ref.~\cite{Baker:1977hp}
(\cite{Orlov:2000wn}), providing
a consistency check on the determination of our coefficients up to ${\cal O}(g^5)$.\footnote{The $v^3$-coefficient of $\eta$ in eq.~(5) of ref.\cite{Orlov:2000wn} is actually in disagreement with ref.\cite{Baker:1977hp} and our results, but we suspect this is simply due to a typo in that formula.} The other coefficients are new.
\begin{table}[t]
\centering
\begin{tabular}{lllc}
\toprule
 Ref., Year \phantom{s}  & \multicolumn{1}{c}{$\nu$\phantom{spppsppppppp}} &   \multicolumn{1}{c}{\hspace{-1.45cm}$\eta$}  &   Method \\
\midrule 
\cite{Onsager:1943jn}, 1944 &  \hspace{1.85cm}$1$ & \hspace{-.85cm}$\frac 14$  &   {\small Exact} \\ &&& \\[-10.5pt]
\cite{Baker:1977hp}, 1978 &   \multirow{2}{*}{\includegraphics[scale=.425]{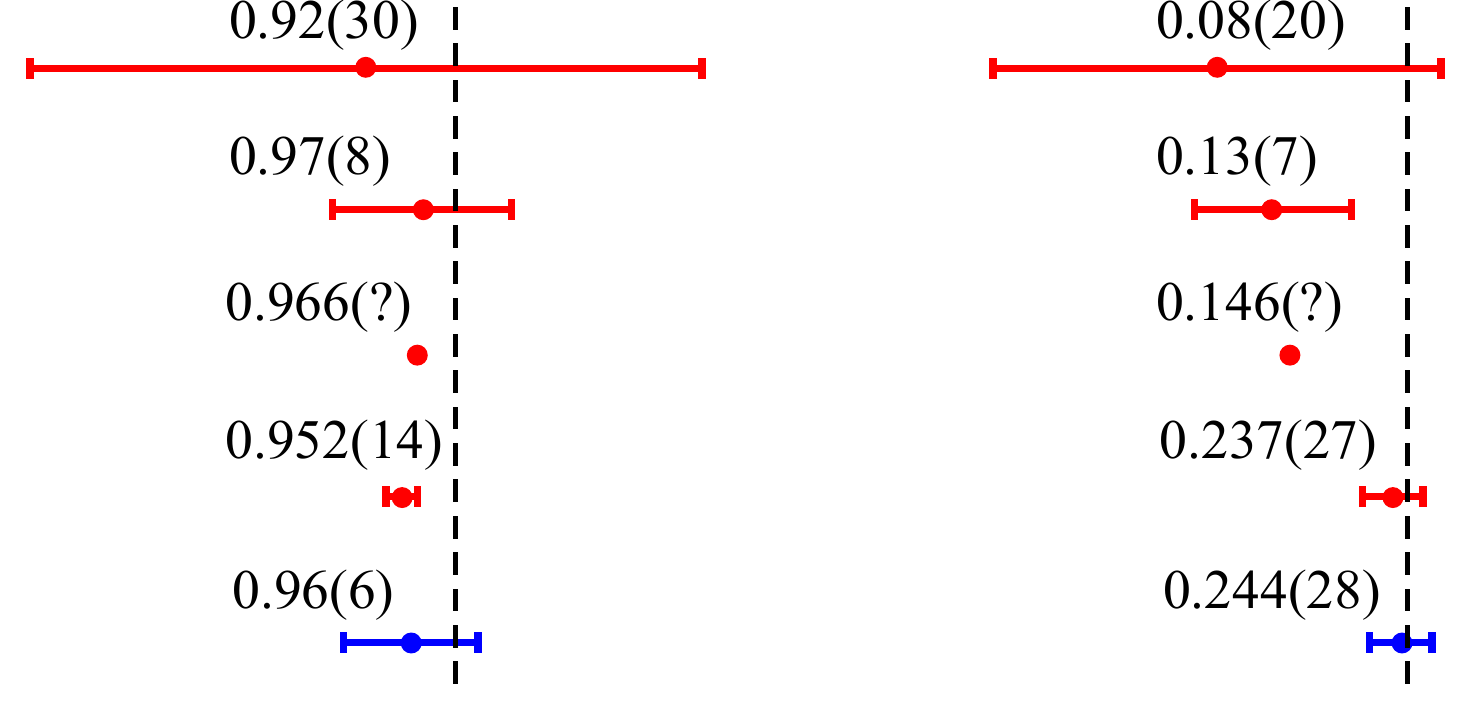}}    &     &  PT$_{\rm 4pt}$, 4 loops \\ & &&\\[-9.5pt]
\cite{LeGuillou:1979ixc}, 1980 & &   &    PT$_{\rm 4pt}$, 4 loops   \\ &&& \\[-9.5pt]
\cite{Orlov:2000wn}, 2000 &  &  &    PT$_{\rm 4pt}$, 5 loops  \\ &&& \\[-9.5pt]
\cite{Kompaniets:2017yct}, 2017 &   &  &    $\epsilon$-expansion, 6 loops  \\ &&& \\[-9.5pt]
This work, 2018 &  &  &   PT$_{\rm 2pt}$, 8 loops ($\nu$), 6 loops $(\eta)$ \\ 
\bottomrule
\end{tabular}
\caption{Comparison of the critical exponents $\nu$ and $\eta$ computed using different resummation techniques at different orders. See the main text for further details.} 
\label{tablenueta}
\end{table}

We compare in table \ref{tablenueta} our results for $\nu$ and $\eta$ with those obtained with PT$_{\rm 4pt}$.
We also include in the table the most recent results in the $\epsilon$-expansion, based on a six loop ($\epsilon^6$) resummation  \cite{Kompaniets:2017yct}. 
Under the column ``Method" we also report the number of loop coefficients used. 
Resummation based on  PT$_{\rm 4pt}$ is known to be not very accurate in $2d$, in contrast to the $3d$ case. 
In particular, as can be seen from table \ref{tablenueta}, the value of $\eta$ significantly differs from its exact value $\eta = 1/4$.
This problem seems to be related to possible non-analyticites in $\beta(g)$ that are not well captured by the Borel resummation \cite{Calabrese:2000dy}. 
This issue is not present in our way of extracting $\eta$ directly from the two-point function, as explained in section \ref{sec:results}. It does not seem to be present in the $\epsilon$-expansion as well.
On the other hand, our estimate for $\nu$ has not improved since ref.~\cite{LeGuillou:1979ixc}. 

Using eqs.~(\ref{betaeta}) and the PT$_{\rm 4pt}$ scheme, we can determine $v^*$ and $\eta=\eta(v^*)$, and compare our results with those in the literature.
Consistency of the coupling expansion requires to keep terms up to ${\cal O}(v^{N+1})$ in $\beta$ and to ${\cal O}(v^{N})$ in $\eta$.
With $N=4$, the number of terms used in refs.\cite{Baker:1977hp,LeGuillou:1979ixc}, we find $v^*=1.84(5)$,  $\eta=0.13(5)$, in very good agreement with
the values  $v^*=1.8(3)$, $\eta=0.08(20)$ in ref.\cite{Baker:1977hp} and $v^*=1.85(10)$, $\eta=0.13(7)$ in ref.\cite{LeGuillou:1979ixc}.
With $N=5$, the number of terms used in ref.\cite{Orlov:2000wn}, we find $v^*=1.82(5)$,  $\eta=0.13(3)$, in agreement with
the values  $v^*=1.837(?)$, $\eta=0.146(?)$ in ref.\cite{Orlov:2000wn}. Taking $N=6$, we find $v^*=1.80(4)$ and $\eta=0.15(3)$, while taking $N=7$ we find $v^*=1.82(4)$ and $\eta=0.16(2)$.  
We see that the accuracy of the results grows very slowly as the order increases. In particular, the value of $\eta$ at $N=7$ is more than 4 standard deviations away 
from $1/4$, confirming the poor accuracy of PT$_{\rm 4pt}$ when applied to the 2d $\phi^4$ theory.

Aside from a numerical comparison, the  PT$_{\rm 2pt}$ developed in our work has some advantages with respect to
 PT$_{\rm 4pt}$. First of all, the proofs in both ref.\cite{Eckmann} and in section \ref{BorelGeom} about the Borel resummability in the 2d $\phi^4$ theory apply for bare, and not renormalized,
quartic coupling $\lambda$. Second, the definition of $g_R$ requires the unavoidable computation of a 4-point function, $\Gamma^{(4)}$, while with
 PT$_{\rm 2pt}$ such computation can be avoided. Typically the higher the Schwinger functions, the less accurate are the results based on numerical resummations. 
Finally, PT$_{\rm 2pt}$ allows a direct comparison with intrinsically non-perturbative methods, such as lattice and Hamiltonian truncations, since their renormalization schemes coincide.

\section{Conclusions}
\label{sec:conclusions}
In this paper we have shown, using arguments based on Lefschetz thimbles in the spirit of refs.\cite{Serone:2017nmd,Serone:2016qog},
that the Schwinger functions in a large class of Euclidean scalar field theories in $d<4$ are Borel reconstructable from their loopwise expansion.
Whenever phase transitions occur, care should be taken in selecting the appropriate vacuum  in the other phase of the theory.
In the context of the 2d $\phi^4$ theory,  we have argued that, in absence of a proper vacuum selection by means of an explicit symmetry breaking term, the Schwinger functions resummed from the unbroken phase might have a different analytic structure in the coupling constant $g$ and the expected singularities associated to the phase transition at $g=g_c$ might not be visible.
The resummed $n$-point functions for $g>g_c$ should be interpreted as $n$-point functions in the $\mathbb Z_2$ broken phase around a vacuum where cluster decomposition is violated. 

Postponing the study of the $\mathbb Z_2$ broken phase of the 2d $\phi^4$ theory to ref.\cite{Z2Broken}, we have focused in this paper on a detailed study of 
the $0$- and $2$-point functions of the 2d $\phi^4$ theory in the unbroken phase with $0\leq g \leq g_c$ and  $m^2>0$. 
We have computed the perturbative series expansion for the vacuum energy and the physical mass up to order $g^8$ and 
Borel resummed the truncated series using known resummation techniques. In this way, for the first time, we accessed the strong coupling behavior
of the 2d $\phi^4$ theory away from criticality using Borel-resummed perturbative expansions. The renormalization scheme chosen allows us a direct comparison with other non-perturbative techniques, most notably Hamiltonian truncation methods. The overall very good agreement of the results can be seen as convincing numerical evidence of the Borel summability of the theory and of the effectiveness of the method.

Given the generality of our arguments in section~\ref{BorelGeom}, our techniques should apply equally
well to many other theories. Natural candidates to consider next include the 3d $\phi^4$ theory and the $O(N)$ 3d vector models.  

\section*{Acknowledgments}

We thank J. Elias-Mir\`o for sharing with us the data needed to compare our results with those in refs.~\cite{Elias-Miro:2017xxf,Elias-Miro:2017tup}  in figs.~\ref{fig:LambdaCMLoops} and \ref{fig:MassCMLoops}.
We thank P. Calabrese, J. Elias-Mir\`o, G. Mussardo, A. Pelissetto and S. Rychkov for useful discussions. 
We acknowledge SISSA and ICTP for granting access at the Ulysses HPC Linux Cluster, and the HPC Collaboration 
Agreement between ICTP and CINECA for granting access to the A3 partition of the Marconi Lenovo system.
Preliminary results of this work have been presented by M.S. at the IHES workshop 
``Hamiltonian methods in strongly coupled Quantum Field Theory", January 8-12, 2018. We thank all the participants to the workshop for useful feedback. 
M.S. acknowledges support from the Simons Collaboration on the Non-perturbative Bootstrap. 

\appendix

\begin{table}[t]
\centering
{
\renewcommand{\arraystretch}{1.2}
\begin{tabular}{c|cccc}
\toprule
k &  $b^{(0)}_k$ & $b^{(1)}_k$ & $b^{(2)}_k$ & $b^{(3)}_k$\\
\midrule
2 & $-3/2$ & $0.0809453264$ & $-0.0128046736$ & $0.0035065405$ \\
3 & $\frac{9}{\pi} + \frac{63 \zeta(3)}{2 \pi^3}$ & $-0.341795194(75)$ & $0.079771437(20)$ \\
4 & $-14.777287(22)$  & $ 1.8559406(86)$ & $-0.5258941(27)$\\
5 & $66.81651(43)$  & $-10.83118(19)$ \\
6 & $-353.2405(28)$ & $68.3310(29)$ \\ 
7 & $2111.715(36)$ \\
8 & $-13994.24(54)$\\
\bottomrule
\end{tabular}
}
\caption{Values of the coefficients $b_k^{(n)}$ for the series expansion of the $n$th-derivative of the two-point function $\widetilde \Gamma _2 ^{(n)}(-m^2)$ as defined in \eqref{eq:Gamma2n-expansion}. These coefficients are the ones needed to get the series of the physical mass $M^2$ up to ${\cal O}(g^8)$ reported in eq.~\eqref{mphFullSeries}. The coefficients $b^{(1)}_2$, $b^{(2)}_2$, $b^{(3)}_2$ are determined numerically with arbitrary precision.}
\label{tab:bkn}
\end{table}
\begin{table}[t]
\centering
\begin{tabular}{c|lllll}
\toprule
$|x|$ & \multicolumn{1}{c}{$g^2$} & \multicolumn{1}{c}{$g^3$} & \multicolumn{1}{c}{$g^4$} & \multicolumn{1}{c}{$g^5$} & \multicolumn{1}{c}{$g^6$}\\
\midrule
$0.005$ & $0.10176449(22)$ & $-0.2637639(86)$ & $0.948600(25)$ & 
$-4.06456(43)$ & $20.0963(51)$\\ 
$0.010$ & $0.10175903(21)$ & $-0.2637468(84)$ & $0.948552(29)$ & 
$-4.06308(71)$ & $20.1215(38)$\\ 
$0.020$ & $0.10173618(21)$ & $-0.2636508(84)$ & $0.948322(26)$ & 
$-4.06241(72)$ & $20.0948(49)$\\ 
$0.050$ & $0.10157963(21)$ & $-0.2631310(29)$ & $0.946740(29)$ & 
$-4.05581(70)$ & $20.0892(40)$\\ 
$0.100$ & $0.10104318(21)$ & $-0.2615325(29)$ & $0.942048(29)$ & 
$-4.03942(76)$ & $20.0100(39)$\\ 
$0.200$ & $0.09908619(21)$ & $-0.2562453(29)$ & $0.926197(25)$ & 
$-3.98035(64)$ & $19.7198(38)$\\ 
$0.300$ & $0.09620711(21)$ & $-0.2488461(28)$ & $0.903417(23)$ & 
$-3.88874(53)$ & $19.3021(38)$\\ 
$0.500$ & $0.08868032(20)$ & $-0.2299485(26)$ & $0.843591(23)$ & 
$-3.64994(65)$ & $18.1834(37)$\\ 
\bottomrule
\end{tabular}
\caption{Series coefficients for the two point function $\langle \phi(x) \phi(0) \rangle$ up to order $g^6$ for some selected values of $x$. We omit here the tree level term--given by $G_0(x)$ defined in eq.~\eqref{eq:G0}--and the $\mathcal{O}(g)$ term which is identically zero in the chosen scheme.}
\label{tab:G2x}
\end{table}

\begin{table}[t]
\centering
\begin{tabular}{c|ccc}
\toprule
order & $\widetilde \Gamma_2(p^2=0)$ & $\partial_{p^2}\widetilde \Gamma_2(p^2=0)$ & $\widetilde \Gamma_4 (\lbrace p_i =0\rbrace)$\\
\midrule
0 & $1$ & $1$ & $0$\\ 
1 & $0$ & $0$ & $24$\\ 
2 & $-\frac{12}{\pi ^{3/2}} G_{3,3}^{3,2}\left(
4\left|
\begin{array}{c}
 0,0,\frac{1}{2} \\
 0,0,0 \\
\end{array}
\right.\right)$ & 
$ \frac{12}{\pi ^{3/2}} G_{3,3}^{3,2}\left(4\left|
\begin{array}{c}
 -1,-1,\frac{1}{2} \\
 0,0,0 \\
\end{array}
\right.\right)$ 
& $-216/\pi$\\ 
3 & $ 3.7798975113$ & $-0.27412237255$ & $270.8452888$\\  
4 & $-13.1529123(81)$ & $1.4204875(19)$ & $-1403.66817(58)$ \\ 
5 & $57.50923(15)$ & $-7.983133(18)$ & $8341.758(61)$ \\ 
6 & $-295.3633(20)$ & $48.89365(40)$ & $-54808.87(32)$ \\ 
7 & $1723.533(29)$ & $-324.0971(70)$ & $392070.3(6.4)$ \\ 
8 & $-11200.30(46)$ & $2312.97(19)$ & $-3.02573(45)\cdot 10^6$ \\
\bottomrule
\end{tabular}
\caption{Series coefficients for the two point function $\widetilde \Gamma_2$, its derivative $\partial_{p^2} \widetilde \Gamma_2$ and the four point function $ \widetilde \Gamma_4 $ at vanishing external momenta.  At order 2 the coefficients of $\widetilde \Gamma_2$ and $\partial_{p^2} \widetilde \Gamma_2$ are expressed in terms of Meijer G-functions. At order 3 the coefficients are determined numerically with arbitrary precision.}
\label{tab:G20-G40}
\end{table}

\section{Coefficients of $2$ and $4$ Point Functions}
\label{sec:appendix}
In this appendix we report the coefficients for the series expansion of the $2$ and $4$ point functions obtained by integration of the Feynman diagrams as explained in section~\ref{sec:pert-coeff}. 

In Tab.~\ref{tab:bkn} we list the coefficients $b_k^{(n)}$ of the $n$th-derivative of the $2$-point function $\widetilde \Gamma _2 ^{(n)}$ at momentum $p^2=-m^2$ %
that are relevant for the determination of the pole mass $M$ up to $g^8$ order. 
Plugging the perturbative expansions of the derivatives $\widetilde \Gamma _2 ^{(n)}(-m^2)$ in eq.~\eqref{Gamma2m2} and solving the equation order by order in $g$ we get the series for $M^2$ as reported in eq.~\eqref{mphFullSeries}.

In Tab.~\ref{tab:G2x} we list the coefficients for the two point function in configuration space $\langle \phi(x) \phi(0) \rangle$ up to order $g^6$ for some selected values of $x$. These coefficients have been used in subsection~\ref{subsec:criticalreg} to determine the critical exponent $\eta$.
We consider values of $x$ within the range $0.005\leq x \leq 0.5$. 
The upper limit $x \leq 0.5$ is given by imposing $M(\Delta g) x\ll 1$, where $\Delta g$ is the uncertainty in the determination of the critical coupling $g_c$. This condition is necessary for the residual exponential factor $\exp(-M(\Delta g) x)$ not to spoil the determination of $\eta$. The lower limit  $x \geq 0.005$ is instead determined by the numerical accuracy we reach for the coefficients and the requirement that the coefficients must be statistically different between each other: this sets a lower bound on the possible $\Delta x$ between two points which is numerically determined as $\Delta x \gtrsim 0.005 $.

In Tab.~\ref{tab:G20-G40} we list the coefficients for the $2$-point function $\widetilde \Gamma_2(p^2=0)$, its derivative $\partial_{p^2} \widetilde \Gamma_2(p^2=0)$ and the four point function $ \widetilde \Gamma_4 (\lbrace p_i =0\rbrace)$ that are needed in the determination of the series for $\beta(g_R)$ and $\eta(g_R)$ as reported in eq.~\eqref{betaeta}.

\end{document}